\documentclass{article}
\usepackage{arxiv}

\usepackage[utf8]{inputenc} 
\usepackage[T1]{fontenc}    
\usepackage{hyperref}       
\usepackage{url}            
\usepackage{booktabs}       
\usepackage{amsfonts}       
\usepackage{nicefrac}       
\usepackage{microtype}      
\usepackage{amsmath}
\usepackage{mathtools}
\usepackage{algorithm}
\usepackage{algorithmic}
\usepackage{stfloats}
\usepackage{url}
\usepackage{multirow}

\usepackage{subcaption}
\usepackage{xcolor,soul}

\DeclareMathOperator*{\argmin}{\underset{\theta} \arg\!\min}
\DeclareMathOperator*{\argminC}{\underset{\mathbf{C}} \arg\!\min}

\title{Examining the Mapping Functions of Denoising Autoencoders in Singing Voice Separation}

\author{
  Stylianos Ioannis Mimilakis\thanks{Corresponding author: \texttt{mis@idmt.fraunhofer.de}} \\
  Fraunhofer-IDMT\\
  Ilmenau, Germany \\
  \And
  Konstantinos Drossos \\
  Audio Research Group \\
  Tampere University\\
  Tampere, Finland
  \And
  Estefan\'{i}a Cano \\
  Fraunhofer-IDMT\\
  Ilmenau, Germany
  \And
  Gerald Schuller \\
  Dpt. for Media Technology\\
  Techinical University of Ilmenau\\
  Ilmenau, Germany
  }

\begin{document}
\maketitle

\begin{abstract}
The goal of this work is to investigate what singing voice separation approaches based on neural networks learn from the data. We examine the mapping functions of neural networks  based on the denoising autoencoder (DAE) model that are conditioned on the mixture magnitude spectra. To approximate the mapping functions, we propose an algorithm inspired by the knowledge distillation, denoted  the neural couplings algorithm (NCA).
The NCA yields a matrix that expresses the mapping of the mixture to the target source magnitude information. Using the NCA, we examine the mapping functions of three fundamental DAE-based models in music source separation; one with single-layer encoder and decoder, one with multi-layer encoder and single-layer decoder, and one using  skip-filtering connections (SF) with a single-layer encoding and decoding. We first train these models with realistic data to estimate the singing voice magnitude spectra from the corresponding mixture. We then use the optimized models and test spectral data as input to the NCA. Our experimental findings show that approaches based on the DAE model learn scalar filtering operators, exhibiting a predominant diagonal structure in their corresponding mapping functions, limiting the exploitation of inter-frequency structure of music data. In contrast, skip-filtering connections are shown to assist the DAE model in learning filtering operators that exploit richer inter-frequency structures.
\end{abstract}

\keywords{Music source separation, singing voice, denoising autoencoder, DAE, skip connections, neural couplings algorithm, NCA.}

\section{Introduction}
\label{sec:intro}
Signal enhancement and separation based on deep learning methods is an active research area that has attracted a lot of attention~\cite{dl_wang_18}. The objective is to estimate an individual target signal from an observed corrupted version. In the context of music source separation, the corrupted observation refers to the observed mixture signal, and the individual target signal to the isolated music source, e.g. singing voice, drums, etc. A particularly challenging task in music source separation is the estimation of singing voice from a single channel, i.e., \textit{monaural}, mixture signal~\cite{rafii18}. To that aim, deep learning approaches are shown to yield state-of-the-art (SOTA) results~\cite{sisec18}.

The majority of deep learning approaches use the time-frequency representation of the mixture signal as input~\cite{sisec18, rafii18}. However, depending on the target output signal they use, we can identify three different classes of approaches. The approaches in the first class, referred to in this text as \textit{spectral approximation} methods, use  the time-frequency representation of the target source as  target output. This input-output relationship of signals follows the seminal work of the DAE model presented in~\cite{vincent_08_den, vincent_den}. The DAE is presented as a model for signal recovery from corrupted observations\footnote{The reader should note that the DAE model, as originally defined in ~\cite{vincent_08_den, vincent_den}, is not restricted to symmetric encoding-decoding functions with dimensionality reduction, commonly referred to as bottleneck~\cite{vincent_08_den}[Sec. 5]. Following this definition, the current article uses the term DAE in a broad sense to refer to methods for signal recovery from corrupted signals.}. 

In the second class of approaches, the target output signal is the pre-computed and source-dependent time-varying filter, i.e., the {\em mask}, that is used for filtering the mixture input signal~\cite{grais17,williamson, mask_targets_speech}. The methods in this category will be referred to in the text as \textit{mask prediction} methods. The usage of pre-computed masks
requires the information of at least two sources, the target and the interfering source(s) contained in the mixture~\cite{grais17, williamson}. That is different from the first class of approaches that require only the time-frequency representation of the target source~\cite{uhl15, grais16}. Furthermore, the computation of masks imposes various assumptions regarding the additive properties of the sources~\cite{liutkus_alpha, voran17}. In some cases, this leads to additional model retraining procedures such that it generalizes to more mixture signals~\cite{grais17}. This has led to research advocating the need for masks to be subject to optimization~\cite{fitz_masks}. 

Although the  mask prediction approaches are not included in the analysis presented in Section \ref{sec:dae_in_mss}, their relevance for the current study comes from the fact that they inspired the third class of approaches. Approaches in the third class conceptually  combine  spectral approximation and mask prediction, and are referred to in the text as \textit{skip-filtering connections}. Specifically, these methods allow deep learning approaches to implicitly mask the input mixture signal by using the output of the deep learning model~\cite{wening14, huang, mim17_mlsp, jannson17}. This operation yields a filtered version of the mixture signal that serves as an estimate of the target source signal, and is used to optimize the overall approach~\cite{wening14, mim17_mlsp, jannson17}. The optimization is performed using a signal reconstruction objective as done in the spectral approximation approaches. In contrast to mask prediction methods~\cite{wening14, mim17, jannson17, drossos18}, methods based on skip-filtering connections do not require pre-computed masks. While the terminology of skip-filtering connections is used in the context of music source separation~\cite{mim17_mlsp, mim17, drossos18}, the speech enhancement and separation community often refers to methods in the third class as the signal approximation method~\cite{wening14, joint_pse}. 

To estimate the target source signal(s), spectral approximation methods usually rely on an additional post-processing step using the generalized Wiener filtering ~\cite{uhl15, uhl17, takahashi17, nug16, grais17, rafii18, mim17, sisec18}. This additional post-processing is an empirical strategy to obtain target signals of better perceptual quality than the direct outputs of the DAEs~\cite{uhl15, uhl17}. In contrast, approaches based on skip-filtering connections, which implicitly mask the mixture signal, yield competitive, yet not superior, results compared to the spectral approximation approaches, without the need of post-processing with Wiener filtering~\cite{drossos18, jannson17}. Furthermore, approaches based on skip-filtering connections tend to result in better separation performance than mask prediction approaches~\cite{mim17_mlsp, mim17}. In relevant literature, for instance in the  large-scale study presented in~\cite{sisec18}, it is not clear why approaches that focus on spectral approximation of the target source ~\cite{uhl15,nug16,takahashi17} require the post-processing step of generalized Wiener filtering. It is also intriguing to provide an explanation on why methods based on skip-filtering connections~\cite{mim17, wening14, jannson17} work well in practice. With this in mind, we formulate the first research question of our work: \emph{\textbf{RQ1 - Why is masking important in approaches based on the DAE model}}?

Additionally, the work presented in~\cite{conservative_autoencoders} underlines the tendency of the encoding and decoding functions of the DAE to become symmetric during training. The composition of symmetric functions yields another function used to map the input to the target signal, i.e., the {\em mapping function}, that shares many similarities with the identity function. For spectral-based denoising, this is a \emph{trivial} scaling of the mixture spectral content. That could potentially result in poor estimation of the target source spectra, unless the learned function is derived from an ideal and time-variant frequency mask~\cite{den_ss}, computed using an appropriate time-frequency masking technique~\cite{ps_masks}. Subsequently, the second research question of our work is: \emph{\textbf{RQ2 - Do DAEs that are commonly employed in music source separation learn trivial solutions for the given problem}}? 

In this study we focus on models that operate on the magnitude spectra of the observed audio mixture, and try to answer these research questions. To that aim we examine the mapping functions of music source separation approaches proposed to estimate the magnitude spectra of the singing voice. Since nearly all music separation approaches are non-linear, the computation of the mapping function is not straightforward. To tackle that, we propose an experimentally derived algorithm that approximates the mapping function of the non-linear model previously optimized for source separation. The result of the algorithm is a matrix that is utilized to linearly map the magnitude information of the mixture to the target source magnitude spectra. We will henceforth denote the algorithm as the {\em neural couplings algorithm} (NCA).

The goal of the NCA is to compute a linear mapping function that describes how the input data are transformed to obtain the desired target source, according to the approach under examination. The NCA differs from methods that aim at explaining the neural network decisions, like for instance the layer-wise relevance propagation method presented in~\cite{LRP_dsp}, but shares many similarities with the optimal transportation theory, in the discrete case~\cite{kolouri_ot_spmag, flamary16_sot}, and the knowledge distillation concept presented in~\cite{hinton_distill}. The conceptual difference between the NCA and the previously stated methods is that the NCA specifically approximates the mapping function of a pre-trained neural network model. It does not aim at compressing the neural network model as in~\cite{hinton_distill}, or computing only distance-dependent mapping(s) between spectral data distributions as in~\cite{flamary16_sot}, or at pin-pointing input spectral features that affect the choice of the neural network model as in~\cite{LRP_dsp}.

In this study, we cluster the spectral approximation and  the skip-filtering connection methods for music source separation  under the general family of DAEs, since these approaches follow the exact same principle as the DAE model presented in~\cite{vincent_08_den, vincent_den}; that is, to recover a target signal from its corrupted version. Specifically, we focus on three particular and fundamental extensions  of the DAE model for music source separation: 
\begin{enumerate}
\item \textbf{DAE}: The DAE model presented in~\cite{vincent_08_den}, as it forms the baseline that source separation approaches have built upon. 
\item \textbf{MSS-DAE}: The multi-layered extension of the DAE following the pioneering works in~\cite{uhl15, nug16}.
\item \textbf{SF}: The implicit mask prediction via the skip-filtering (SF) connections employed in~\cite{mim17_mlsp,wening14,jannson17}. 
\end{enumerate}

In the three models, we employ the rectified linear unit activation function (ReLU) as it was shown experimentally  to perform well in music source separation tasks~\cite{uhl15}. The target signal to be estimated by each model is the singing voice magnitude spectra. For assessing the mapping functions of each model, we use the outcome of the NCA, and objectively compute a fraction of the magnitude contained in the main and off diagonal elements of the mapping matrix computed by the NCA. This will be explained in details in the following sections. 

The rest of this document is organized as follows: Section~\ref{sec:dae_in_mss} provides background information on the DAEs and the variations proposed for the problem of music source separation tasks. Our proposed NCA algorithm is described in Section~\ref{sec:proposed_method}, followed by
 the experimental procedure described in Section~\ref{sec:experiments}. Our experimental findings are presented and discussed in Section~\ref{sec:results}. Section~\ref{sec:conclusions} concludes this work and suggests future research directions.

%
%
\section{Denoising Autoencoders in Music Source Separation}
\label{sec:dae_in_mss}
%
%
\subsection{Background}
\label{subsec:dae_in_mss_A}
Music source separation based on DAEs relies on a supervised learning scenario. Formally, given a data-set $\mathcal{D} = \{\tilde{x}^{(i)}, x^{(i)}\}^{K}_{i=1}$, comprised of $K \in \mathbb{Z}^{+}$ training examples indexed by $i$, the goal is to learn a denoising function $f$. The function $f$ is parameterized by $\theta$, and estimates the clean $x$ from the noisy $\tilde{x}$ observation, i.e., $f:\theta\times\tilde{x}\mapsto x$. Obtaining $\tilde{x}$ involves a mixing process, which for audio signals, is commonly assumed to be the addition of the interfering or noise signal $x_{\text{n}}$ and the target source signal $x$~\cite{mim17}, i.e., $\tilde{x} = x + x_{\text{n}}$. Given a reconstruction loss function $\mathcal{L}$, the parameters $\theta$ are optimized using
\begin{equation}
	\theta^{\mathit{o}} = \argmin\sum_{i=1}^{K}\mathcal{L}(x^{(i)}, \hat{x}^{(i)}) \text{,}
	\label{eq:1}
\end{equation}
\noindent
where $\hat{x}$ is an estimate of $x$, and $\theta^{\mathit{o}}$ is the (ideally optimal) parameter or set of parameters, that minimizes the cost. The updates of the parameters towards obtaining $\theta^{\mathit{o}}$ is carried out using stochastic gradient descent over samples drawn from the data-set $\mathcal{D}$.

\begin{figure}[!t]
\centering
\begin{minipage}{0.24\columnwidth}
\def\svgwidth{\columnwidth}
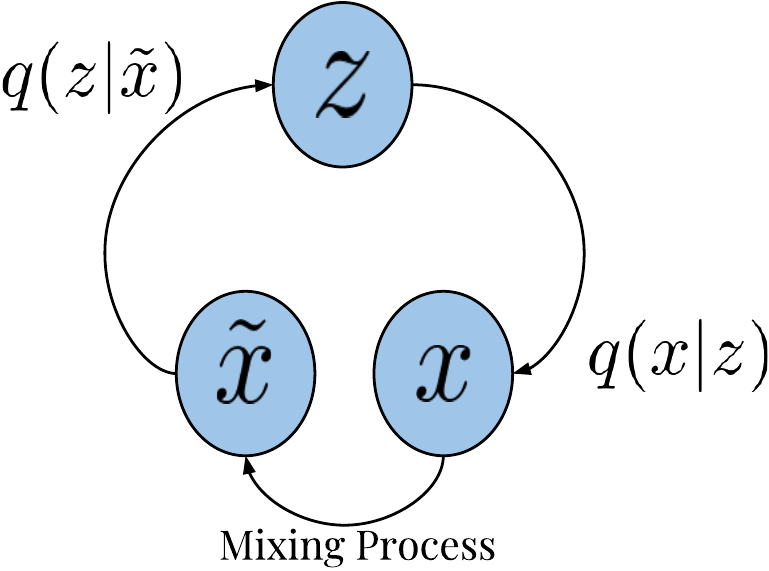
\subcaption{DAE}
\end{minipage}%
\begin{minipage}{0.24\columnwidth}
\def\svgwidth{\columnwidth}
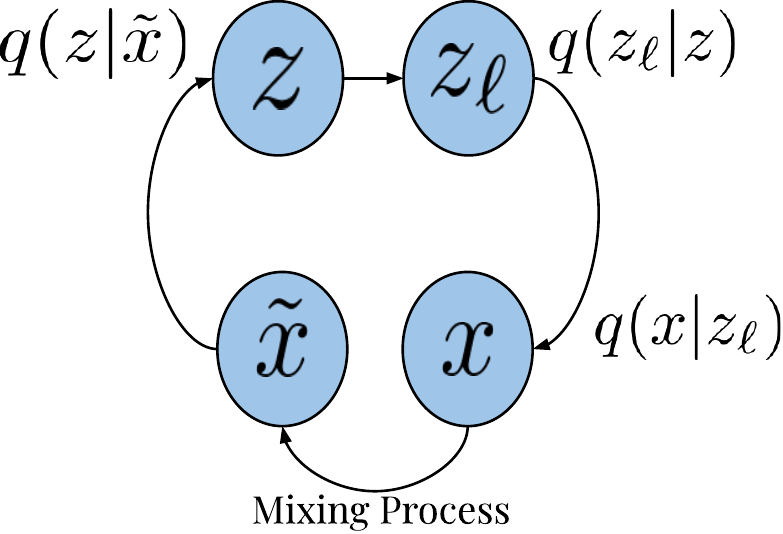
\subcaption{MSS-DAE}
\end{minipage}%
\def\svgwidth{\columnwidth}
\begin{minipage}{0.24\columnwidth}
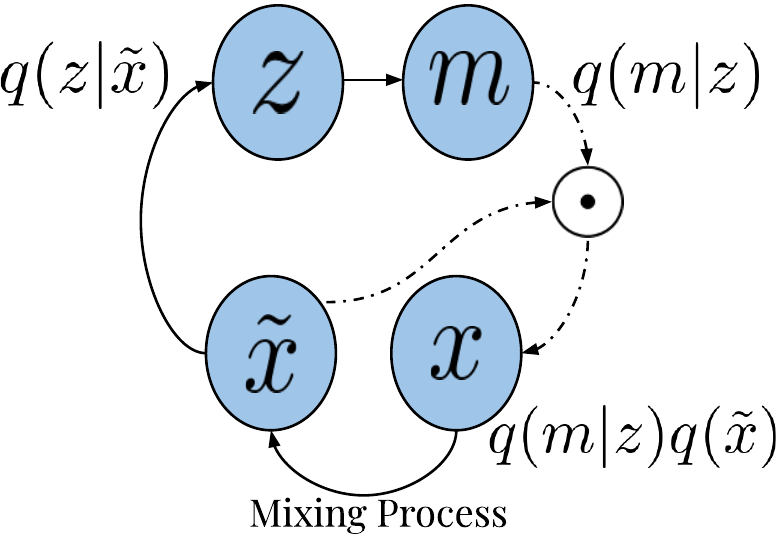
\subcaption{SF}
\end{minipage}
\caption{Illustration of the graphical models of encoder-decoder configurations examined in this work. {\emph{(a) DAE}}: a denoising auto-encoder model~\cite{vincent_08_den}. {\emph{(b) MSS-DAE}}: a three layer example of a DAE model adapted to music source separation~\cite{uhl15,nug16}. {\emph{(c) SF}}: skip-filtering connections~\cite{mim17_mlsp,mim17, wening14, jannson17}. Solid arrows are functions computed by neural networks. Dashed arrows are the identity function. In the context of the input and latent variables, the symbol ``$\odot$'' refers to the multiplication of the corresponding variables.}
\label{fig:pgm}
\end{figure}

In~\cite{vincent_08_den}, two functions are introduced through the DAE structure in order to approximate $f$, namely $f_{\text{enc}}:\theta_{\text{enc}}\times\tilde{x}\mapsto z$ and $f_{\text{dec}}:\theta_{\text{dec}}\times z\mapsto x$. The parameters $\theta := \{\theta_{\text{enc}}, \theta_{\text{dec}}\}$ are optimized with respect to Eq.~\eqref{eq:1}. In the context of music source separation,
the motivation is to learn the empirical distribution $q(x|\tilde{x})$ through the usage of a latent representation $z$, and the utilization of the decoding process $f_{\text{dec}}$. The benefit of incorporating the latent variable $z$ into the model is that it provides a feature space that is useful for denoising auto-encoding~\cite{vincent_08_den} and music source separation~\cite{uhl15, grais17, nug16}. An illustration of the DAE model is given in Fig.~\ref{fig:pgm}a.

In music source separation approaches, the computation of the latent variable $z$ plays an important role~\cite{grais17,uhl17,nug16,uhl15}. Specifically, the performance of the methods and the approximation of the target source $x$ is shown empirically to be based on the computation of $z_{\ell}$, a deeper hidden representations of $z$ that leads to the conditional distribution $q(z_{\ell}|z_{\ell-1})$~\cite{grais17,nug16,uhl15} . The subscript $\ell \in \{1,2,\ldots, L\}$ denotes the depth of the computed, hidden representations. The corresponding graphical depiction of  a three layer example of the MSS-DAE is given in Fig.~\ref{fig:pgm}b.

It is important to note that source separation approaches based on skip-filtering connections use the same two functions as  the DAE, i.e.,
$f_{\text{enc}}$ and $f_{\text{dec}}$. The difference is that $f_{\text{dec}}:\theta_{\text{dec}}\times z\mapsto m$, and
\begin{equation}
x = m \odot \tilde{x}\text{, }
\label{eq:psi_func}
\end{equation}
\noindent
where $m$ is the mask. Eq.~\eqref{eq:psi_func} is implemented by the skip connections which allow $\tilde{x}$ to be propagated to the encoding and to the last decoding function
of the model~\cite{mim17_mlsp, wening14,joint_pse}. SF models the empirical distribution $q(x|\tilde{x})$ using $q(m|\tilde{x})q(\tilde{x})$, as illustrated in~Fig.\ref{fig:pgm}c. Subject to the target source $x$, the  empirical distribution leads to $q(x|\tilde{x})q(\tilde{x})$ since both the mask $m$ and the target signal $x$ are computed as a function of $\tilde{x}$, i.e., $x = f_{\text{dec}}(f_{\text{enc}}(\tilde{x})) \odot \tilde{x}$.
This is conceptually different from the DAE and MSS-DAE, which model $q(x|\tilde{x})$ directly. For the SF model and its corresponding conditional, the product between distributions is the product of the probability values between the outcome of the DAE and $\tilde{x}$.
%
%
\subsection{Prior Work}\label{subsec:sota}
The most widely adopted way to perform music source separation using deep learning is to employ the magnitude spectral representations computed using the short-time Fourier transform (STFT). This is performed in order to reduce the overlap that the sources exhibit in the time-domain signal representation~\cite{giannoulis11}, and to exploit the wide-sense stationarity and the phase-invariant structure(s) of specific types of music sources~\cite{kam14}. Therefore, we can think of the variables of the DAE, MSS-DAE, and SF models as vectors containing magnitude spectral information and allow the symbol ``$\odot$'' to denote the Hadamard (element-wise) product.
However, since the phase information is not considered, the additive properties of the mixing process for computing $\tilde{x}$ do not longer hold, i.e., $\tilde{x} \neq x + x_{\text{n}}$. 
More specifically, 
$\tilde{\mathbf{x}},\,\hat{\mathbf{x}},\text{ }\mathbf{x} \in \mathbb{R}_{\geq 0}^{N}$, 
and $\mathbf{z} \in \mathbb{R}^{F}$ are the mixture signal, the estimated target source signal, the target source signal, and the latent representation after the application of the ReLU function, respectively. $N$ and $F$ denote the dimensionality of the input and hidden representations, respectively.

For source estimation based on DAEs, the authors in~\cite{nug16} propose the use of multi-layered feed-forward neural networks, using context information of past and future STFT magnitude spectra. Context information for spectral-based denoising via feed-forward neural networks is also proposed in~\cite{uhl15}. Aiming to model the dependencies of adjacent time-frames, the work in~\cite{uhl17} proposes to use bi-directional RNNs instead of feed-forward neural networks. In both approaches~\cite{uhl17} and~\cite{nug16}, the estimated, by the DAE, target sources are further processed using the multi-channel Wiener filtering. In contrast, the work presented in~\cite{grais17} proposes to explicitly predict pre-computed time-frequency masks using deep neural networks. After the mask prediction, the masks are applied to the mixture signal and then refine the estimates using DAEs. 
More robust approaches allow implicit mask prediction by introducing the time-frequency masking operation into the computational graphs~\cite{mim17_mlsp, mim17, wening14, huang, wang15, jannson17,nikkunen18} by incorporating the skip connections described in Section~\ref{subsec:dae_in_mss_A}. 

The skip connections are a straightforward extension of the denoising source separation (DSS) framework in the spectral domain presented in~\cite{den_ss}. In the DSS framework, it is proposed to perform spectral-based denoising by learning a sparse matrix with non-zero elements only on the main diagonal. These elements allow a scalar filtering operation of each corresponding frequency sub-band~\cite{wiener_matrix_guy}. In music source separation works based on deep learning, the  frequency sub-band scaling is achieved by employing the Hadamard product. That enables the usage of architectures that encompass time varying information, such as RNNs and/or CNNs, and therefore the prediction of time varying frequency masks.
More specifically, the work in~\cite{huang} proposes to employ deep RNNs for estimating the magnitude of all the sources contained in the mixed signal. The output estimates are then given to a deterministic function which yields source-dependent time-frequency masks by computing the ratio of the estimates. In this approach, RNNs do not learn the masking process, but rather learn to deliver magnitude estimates of the source that can be used to compute the mask. Aiming to also learn the masking process, i.e., allow the neural network to generate mask estimates without the necessity of devising rational models like in~\cite{huang}, the skip-filtering connections were introduced in~\cite{mim17_mlsp}, where an RNN encoder-decoder is responsible for the generation of a mask that estimates the singing voice. Extensions that improve the mask generation process of the model presented in \cite{mim17_mlsp} are discussed in~\cite{mim17, drossos18}. An alternative architecture is presented in~\cite{jannson17}, where ladder-like structured CNNs are proposed for singing voice separation via the aforementioned mask generation process. 
%
%
\subsection{Implementation of the Models}
\label{subsec:model_details}
Focusing on the graphical models presented in Figure~\ref{fig:pgm}, we constrain the problem to fully connected, feed-forward neural network (FNN) layers, and to the minimization of the mean squared error (MSE) loss function, defined as:
\begin{equation}
	\label{eq:mse}
	\mathcal{L}_{\text{MSE}} (\mathbf{x}^{(i)}, \mathbf{\hat{x}}^{(i)}) = \frac{1}{N} ||\mathbf{x}^{(i)} - \mathbf{\hat{x}}^{(i)}||^{2}_{2} \text{,}
\end{equation}
where $||\cdot||_{2}$ denotes the $\ell_2$ vector norm.

Stochastic gradient descent is performed to optimize the model parameters with respect to Eq.~\eqref{eq:mse}. This training configuration including the MSE was adopted from SOTA approaches in music source separation~\cite{takahashi17, uhl17, mim17_mlsp, nug16, uhl15, mim_asilomar}. An example of the calculation of $z$ and the approximation of the $i$-th data example of the source $\hat{\mathbf{x}}$, using the mixture signal $\tilde{\mathbf{x}}$ and the corresponding encoding and decoding functions is given in Eqs.~\eqref{eq:z-dae}--\eqref{eq:relu} for the DAE, MSS-DAE, and SF models.
\begin{align}
    \label{eq:z-dae}
    \mathbf{z}_{\text{DAE}}^{(i)} &= g(\mathbf{W}_{\text{enc}}\tilde{\mathbf{x}}^{(i)} + \mathbf{b}_{\text{enc}}),\\
	\label{eq:dae}
	\mathbf{\hat{x}}_{\text{DAE}}^{(i)} &= g(\mathbf{W}_{\text{dec}}\mathbf{z}_{\text{DAE}}^{(i)} + \mathbf{b}_{\text{dec}})\text{,}\\
	\label{eq:z-mssdae_x}
	\mathbf{z}_{\text{MSS-DAE}}^{(i)} &= g(\mathbf{W}^{(\ell=1)}g(\mathbf{W}_{\text{enc}}\tilde{\mathbf{x}}^{(i)} + \mathbf{b}_{\text{enc}}) + \mathbf{b}^{(\ell=1)})\text{,}\\
	\label{eq:mssdae_x}
	\mathbf{\hat{x}}_{\text{MSS-DAE}}^{(i)} &= g(\mathbf{W}'_{\text{dec}}\mathbf{z}^{(i)}_{\text{MSS-DAE}} + \mathbf{b}'_{\text{dec}})\text{,}\\
	\label{eq:z-sf}
	\mathbf{z}_{\text{SF}}^{(i)} &= g(\mathbf{W}''_{\text{enc}}\tilde{\mathbf{x}}^{(i)} + \mathbf{b}''_{\text{enc}}),\\
	\label{eq:sf}
	\mathbf{\hat{x}}_{\text{SF}}^{(i)} &= g(\mathbf{W}''_{\text{dec}}\mathbf{z}_{\text{SF}}^{(i)} + \mathbf{b}''_{\text{dec}}) \odot \tilde{\mathbf{x}}^{(i)}\text{, where}\\
	\label{eq:relu}
	g(x) &= \text{max}(0, x).
\end{align}
Eqs.~\eqref{eq:z-dae},~\eqref{eq:z-mssdae_x}, and~\eqref{eq:z-sf} describe the encoding functions, and Eqs.~\eqref{eq:dae},~\eqref{eq:mssdae_x}, and~\eqref{eq:sf} the decoding functions of each model. The weight matrices and bias terms are denoted by $\mathbf{W}\text{ and }\mathbf{b}$, respectively, and the subscripts ``enc'' and ``dec'' stand for encoder and decoder layers, respectively. The superscripts ``$'$'' and ``$''$'' in the weights and biases are used to distinguish between the parameters of different models (e.g. between $\mathbf{W}$ of MSS-DAE and DAE models). In more detail, Eqs.~\eqref{eq:z-dae} and~\eqref{eq:dae} express the encoding and decoding functions for the DAE model.
An example of a three layered MSS-DAE model, with a single decoding function, for approximating the target source is given by Eqs.~\eqref{eq:z-mssdae_x} and~\eqref{eq:mssdae_x}. The SF model in Eqs.~\eqref{eq:z-sf} and~\eqref{eq:sf} employs the same encoding and decoding configuration as the DAE, with the only difference the output of the decoding is element-wise multiplied with the input to the model. In the Eqs.~\eqref{eq:z-dae}--\eqref{eq:sf} the encoding and decoding functions are realized as linear operators, i.e., vector-matrix product,  that are then followed by the ReLU activation function expressed in Eq.~\eqref{eq:relu}.
%
%
\section{Neural Couplings Algorithm}
\label{sec:proposed_method}
The NCA is an iterative method for approximating the mapping function
of a non-linear source separation model. The mapping function is defined as an affine transformation of magnitude spectral data. We constrain the affine transformation to be linear and model-dependent in order to allow us to intuitively examine what each source separation model has learned. The affine transformation is represented by a matrix that we denote  the {\em couplings matrix} $\mathbf{C} \in \mathbb{R}^{N \times N}$. The couplings matrix $\mathbf{C}$ is used to transform/map the input mixture magnitude spectrum $\tilde{\mathbf{x}}$ to the output $\mathbf{y} \in \mathbb{R}_{\geq 0}^{N}$ of the last layer of each corresponding model including the non-linearity, i.e., the output of the decoding matrix followed by the ReLU function. The vector $\mathbf{y}$ is used to denote the output of the decoding function in each model. Specifically, the output for the DAE and MSS-DAE models is the singing voice spectra, and for the SF model is the derived frequency mask. The indexing by $i$ in $\tilde{\mathbf{x}}$ and $\mathbf{y}$ is dropped in order to denote the usage of spectral data that are not sampled from the training data-set. 

The reason for using the mask instead of singing voice spectra for the SF model follows naturally from the graphical models illustrated in Fig.~\ref{fig:pgm}. Particularly for the SF model, the information of the mixture spectra is also necessary after the decoding process in order to estimate the source via masking using the Hadamard product expressed in Eq.~(\ref{eq:sf}). As masking is an additional operator that heavily depends on the mixture data, the computation of a single affine transformation would fail to approximate both the masking and the mapping function of the model. A solution to this is given by knowledge distillation~\cite{hinton_distill}; that is, to use the last hierarchical variable that is computed using the model's parameters, leading to fair usage of information during the approximation of the NCA among source separation models.
In our study, we also include the ReLU function applied to the decoding stage since an algebraic expression for the ReLU function is given in~\cite{pascanu_montufar}, and its relevance is explained later in this section.
 
In the ideal case that each model is linear, the couplings matrix, and thus the mapping function, is expressed algebraically as the product of the corresponding encoding and decoding matrices. We denote that product as the {\em linear composition}. Neglecting the bias terms for brevity in the notation, the linear composition is computed for each model as follows:
\begin{align}
	\label{eq:comb_dae}
	\mathbf{C}^{\mathit{o}}_{\text{linear-DAE}} &= \mathbf{W}_{\text{dec}} \mathbf{W}_{\text{enc}}\text{,} \\
	\label{eq:comb_mssdae}
\mathbf{C}^{\mathit{o}}_{\text{linear-MSS-DAE}} &= \mathbf{W}_{\text{dec}}'\mathbf{W}^{\small(\ell=1)}\mathbf{W}_{\text{enc}}'\text{, and} \\
	\label{eq:comb_sf}
\mathbf{C}^{\mathit{o}}_{\text{linear-SF}} &= \mathbf{W}_{\text{dec}}'' \mathbf{W}_{\text{enc}}''\text{.}
\end{align}

As the DAE, MSS-DAE, and SF, models are non-linear, directly employing  Eqs.~\eqref{eq:comb_dae}--\eqref{eq:comb_sf} would result into rather crude approximations of the models' mapping functions. Even the linear behaviour of the ReLU function in the non-negative range, and of the vector-matrix products expressed in Eqs.~\eqref{eq:z-dae}--\eqref{eq:sf}, is not sufficient for the above linear composition functions to hold. The ReLU function performs a thresholding operation on the variables that yield the latent $\mathbf{z}$ and output $\mathbf{y}$ vectors  that are learned through observations drawn from the training data-set. This in turn makes the models highly non-linear~\cite{papyan_ieee}. An algebraic expression of the ReLU function that is related to the concept of thresholding of negative values is presented in~\cite{pascanu_montufar}. In~\cite{pascanu_montufar}[Sec. 3, Eq.~(3)], a single application of the ReLU function for a given input vector $\tilde{\mathbf{x}}$ can be expressed as a {\em binary diagonal} matrix, that sets to $0$ any negative value of the encoded vector. We denote this matrix  $\mathbf{G}$. Consequently, to obtain the couplings matrix of the DAE, MSS-DAE, and SF models, it is necessary to compute as many matrices $\mathbf{G}$ as the number of application of the ReLU function in the DAE, MSS-DAE, and SF models:
\begin{equation}
    \mathbf{C}^{\mathit{o}} = \mathbf{G}_{\text{dec}}\mathbf{W}_{\text{dec}} \ldots \mathbf{G}_{\text{enc}}\mathbf{W}_{\text{enc}}.
    \label{eq:non_linear_comb}
\end{equation}

Furthermore, to compute each matrix $\mathbf{G}_{*}$ in Eq.~\eqref{eq:non_linear_comb}, it is necessary to learn the model specific dependencies that are captured by the DAE, MSS-DAE, and SF models during the supervised training ~\cite{papyan_ieee, pascanu_montufar}.  The asterisk ``$*$'' in the notation is used for brevity, and replaces the subscripts and/or superscripts of the layer identifiers initially used in Eqs.~\eqref{eq:z-dae}--\eqref{eq:sf}. The data dependencies expressed by each $\mathbf{G}_{*}$ refer to the algebraic operations between the mixture $\tilde{\mathbf{x}}$ or the corresponding latent vectors $\mathbf{z}$, i.e., the encoding or decoding matrices $\mathbf{W}_{*}$, and the corresponding bias terms $\mathbf{b}_{*}$ as in Eqs.~\eqref{eq:z-dae}--\eqref{eq:relu}.

A straightforward way to learn the data dependencies can be derived from the knowledge distillation concept presented in~\cite{hinton_distill}. In knowledge distillation, a neural network, i.e., the {\em student}, is optimized by means of (stochastic) gradient descent to predict the output of a more complicated model (e.g. the non-linear DAE, MSS-DAE and SF). Subject to the goal of this work, we employ gradient descent as in knowledge distillation~\cite{hinton_distill}, and we restrict the student network to be a linear, affine transformation from $\tilde{\mathbf{x}}$ to $\mathbf{y}$. That transformation is based on the product of the mixture spectra $\tilde{\mathbf{x}}$ and the couplings matrix $\mathbf{C}$. For computing the couplings matrix $\mathbf{C}$ we propose to solve the following optimization problem: 
\begin{equation}
\label{eq:argmin_c}
\mathbf{C}^{\mathit{o}} = \argminC || \mathbf{y} - \mathbf{C} \tilde{\mathbf{x}}||, {\text{ where}}
\end{equation}
$||\cdot||$ is the $\ell_1$ norm and $\tilde{\mathbf{x}}$ is sampled from the testing data-set. The output vector $\mathbf{y}$ is computed by using $\tilde{\mathbf{x}}$ as an input to the corresponding model. We propose to use the $\ell_1$ instead of the $\ell_2$ norm, employed in the optimization of the source separation models, because the vector $\mathbf{y}$ is expected to be sparse, due to the application of the ReLU function~\cite{papyan_ieee}. Commonly, the $\ell_1$ norm offers an attractive objective for error minimization between sparse vectors, when the choice of the regularization strength parameter in the sparse aware setting of $\ell_2$-based optimization is difficult~\cite{theodoridis_ml}[Ch. 9]. Nonetheless, by minimizing the $\ell_1$ norm of errors we assume that the expected reconstruction error follows an exponential distribution. Given that Eq.~\eqref{eq:argmin_c} is inspired by the student network of the knowledge distillation concept~\cite{hinton_distill}, we will denote this strategy as the {\em student}.

To compute $\mathbf{C}^{\mathit{o}}$, the student strategy employs the following partial derivatives, with $E$ denoting the error of the $\ell_1$ norm contained in Eq.~(\ref{eq:argmin_c}):
\begin{equation}
\label{eq:dist_pd}
\mathbf{\Delta} :=
\frac{\partial E}{\partial \mathbf{C}} = \frac{\partial E}{\partial \mathbf{C}\tilde{\mathbf{x}}}\frac{\partial \mathbf{C}\tilde{\mathbf{x}}}{\partial \mathbf{C}}
\end{equation}
From Eq.~\eqref{eq:dist_pd}, it follows that the gradient signal $\mathbf{\Delta}$ that is used to update $\mathbf{C}$ in an iterative manner is given by $\mathbf{\Delta} = \text{sgn}(\mathbf{C}\tilde{\mathbf{x}} - \mathbf{y}) \tilde{\mathbf{x}}^T$, where $\text{sgn}$ is the signum element-wise function and $\cdot^T$ is the vector/matrix transposition. The gradient signal $\mathbf{\Delta}$ suggests that the optimal $\mathbf{C}^{\mathit{o}}$
lies over the least affinity between the mixture magnitude spectra $\tilde{\mathbf{x}}$ and $\mathbf{C}\tilde{\mathbf{x}} - \mathbf{y}$.
This means that the updates of $\mathbf{C}$ only favor the minimization of the  reconstruction error term. Although this strategy could yield simplified and robust surrogates of possibly deep and complex models for source separation (in terms of reconstruction errors)~\cite{hinton_distill}, it neglects the learned data dependencies contained in the encoding and decoding matrices of Eq.~\eqref{eq:non_linear_comb}.

According to~\cite{wang16_datajacobian}, the learned data dependencies of the encoding and decoding functions are the key ingredient to characterize the functionality of a non-linear model. It is also shown in~\cite{wang16_datajacobian} that those dependencies are described by linear systems that can be computed using observations of $\tilde{\mathbf{x}}$, $\mathbf{y}$ and the corresponding weight matrices. Given that our goal is to approximate the functionality of the model, i.e., including the knowledge captured by the encoding/decoding matrices $\mathbf{W}_{*}$ and the bias terms $\mathbf{b}_{*}$, we propose an alternative strategy that we denote as {\em compositional}. The proposed strategy composes the couplings matrix $\mathbf{C}$
similar to the composition expressed in Eq.~\eqref{eq:non_linear_comb}. In contrast to the method presented in~\cite{wang16_datajacobian}, the compositional approach does not require the explicit computation of the latent information $\mathbf{z}_{*}$ of each model. Instead, it uses directly $\mathbf{G}_{*}$ to extract data dependencies contained in each $\mathbf{W}_{*}$, capturing both the information of the encoding/decoding matrices and the spectral data. More specifically, we exploit the fact that the ReLU function behaves linearly in the non-negative range where the mixture $\tilde{\mathbf{x}}$, the output $\mathbf{y}$, and the latent $\mathbf{z}$ vectors reside in. Therefore, and according to Eqs.~\eqref{eq:z-dae}--\eqref{eq:sf}, the relevant components that affect which elements are thresholded by the ReLU function for computing the latent $\mathbf{z}$ and output $\mathbf{y}$ vectors, are the row-vectors contained in each $\mathbf{W}_{*}$ and the corresponding bias term(s) $\mathbf{b}_{*}$.

Inspired by~\cite{pascanu_montufar}, that models the ReLU function as a diagonal matrix $\mathbf{G}_{*}$ that scales the row-vectors of $\mathbf{W}_{*}$ in Eq.~\eqref{eq:non_linear_comb}, we build the couplings matrix for the compositional strategy as follows: 
\begin{equation}
\label{eq:cond_a}
\mathbf{C} = (\mathbf{G}_{\text{dec}}\odot\mathbf{W}_{\text{dec}})\ldots(\mathbf{G}_{\text{enc}}\odot\mathbf{W}_{\text{enc}}).
\end{equation}
The Hadamard product in Eq.~\eqref{eq:cond_a} accounts 
for each individual element of the row-vectors contained in the corresponding $\mathbf{W}_{*}$, rather than having a single scalar value per row-vector as in the case of the matrix product using the diagonal matrix. Practically, this mitigates the usage of binary diagonal matrices allowing more degrees of freedom into the approximation as the operator is applied to all the elements of the corresponding matrix vectors~\cite{pascanu_montufar}. The reason for accounting for all the elements, is that the sign and the magnitude of each corresponding element in each $\mathbf{W}_{*}$ carry the essential information in the model-dependent processing of non-negative vectors. To consider the influence of the sign of the elements in each $\mathbf{W}_{*}$, we restrict each matrix $\mathbf{G}_{*}$ to be non-negative, i.e., $\mathbf{G}_{*}\in \mathbb{R}_{\geq 0}^{N \times N}$. To do so, and to account for the influence of the bias terms we compute each $\mathbf{G}_{*}$ using:
\begin{align}
\mathbf{G}_{*} &= g(\hat{\mathbf{G}}_{*}) \\
\hat{\mathbf{G}}_{*} &= \mathbf{P}_{*}(\mathbf{W}_{*} + \mathbf{b}_{*})^T .
\label{eq:cond_b}
\end{align}
In~Eq. \eqref{eq:cond_b} we use a less conventional notation to denote the matrix-vector addition. Given a matrix $\mathbf{W}$ and a vector $\mathbf{b}$ we define the matrix-vector addition as $\hat{W}_{\alpha, \beta} = W_{\alpha, \beta} + b_{\beta}$ for all elements $\alpha$ and $\beta$ that are available in $\mathbf{W}$ and $\mathbf{b}$.
Specifically for Eq. \eqref{eq:cond_b}, we add the bias vector $\mathbf{b}_{*}$ to each column-vector of $\mathbf{W}_{*}$. By the application of $\mathbf{G}_{*}$ via the Hadamard product, the elements in $\mathbf{W}_{*}$ are either preserved and scaled or nullified. This depends on their relevance for mapping the mixture magnitude spectra under linear constraints. The linear constraints are imposed by using the matrix $\mathbf{C}$ to the optimization problem defined in~Eq.~\eqref{eq:argmin_c}. In order to compute the previously mentioned relevance, we propose to learn an additional affine transformation computed for each matrix $\mathbf{W}_{*}$ plus the corresponding bias term $\mathbf{b}_{*}$. This affine transformation is denoted by the matrix $\mathbf{P}_{*}$. The corresponding basis vectors of $\mathbf{P}_{*}$ are unknown for the compositional strategy. To jointly compute the unknowns of the compositional strategy, we use the back-propagation method.
 
 To further analyze how the previously mentioned data dependencies are learned using the compositional strategy, let us consider the case in which a single encoding and decoding matrix is used, as in the DAE and SF models. For the compositional strategy, it is necessary to compute two matrices $\mathbf{P}_{\text{enc}}$ and $\mathbf{P}_{\text{dec}}$. The corresponding partial derivatives are defined as $\frac{\partial E}{\partial \mathbf{P}_{\text{dec}}} = \frac{\partial E}{\partial \mathbf{C}} 
\frac{\partial \mathbf{C}}{\partial \mathbf{G}_{\text{dec}}} \frac{\partial \mathbf{G}_{\text{dec}}}{\partial \mathbf{P}_{\text{dec}}}$, and $\frac{\partial E}{\partial \mathbf{P}_{\text{enc}}} = \frac{\partial E}{\partial \mathbf{C}} \frac{\partial \mathbf{C}}{\partial \mathbf{G}_{\text{enc}}} \frac{\partial \mathbf{G}_{\text{enc}}}{\partial \mathbf{P}_{\text{enc}}}$, respectively. 
Using Eq.(\ref{eq:dist_pd}) the aforementioned partial derivatives are as follows:
\begin{align}
    \label{eq:rest_learn_dec}
    \frac{\partial E}{\partial \mathbf{P}_{\text{dec}}}&= (\mathbf{\Delta}(\mathbf{W}_{\text{enc}} \odot \mathbf{G}_{\text{enc}}) \odot \mathbf{W}_{\text{dec}}  \odot g'(\hat{\mathbf{G}}_{\text{dec}}))(\mathbf{W}_{\text{dec}} + \mathbf{b}_{\text{dec}}) \\
    \frac{\partial E}{\partial \mathbf{P}_{\text{enc}}}&= ((\mathbf{W}_{\text{dec}} \odot \mathbf{G}_{\text{dec}})^T\mathbf{\Delta} \odot \mathbf{W}_{\text{enc}}  \odot g'(\hat{\mathbf{G}}_{\text{enc}}))(\mathbf{W}_{\text{enc}} + \mathbf{b}_{\text{enc}}),
    \label{eq:rest_learn_enc}
\end{align}
where $g'(x)$ is the first derivative of the ReLU element-wise function, that is approximated by a function that is equal to $1$ for positive inputs and $0$ otherwise. 
$\mathbf{G}_{\text{dec}}$ and $\mathbf{G}_{\text{enc}}$ are computed using Eq.~(\ref{eq:cond_b}). In Eqs.~\eqref{eq:rest_learn_dec} and~\eqref{eq:rest_learn_enc},
instead of considering only the gradient for minimizing the reconstruction error $\mathbf{\Delta}$, the models' optimized parameters partake into the optimization of the compositional strategy. Specifically, $\mathbf{W}_{\text{*}}\text{ and }\mathbf{b}_{\text{*}}$ contribute to the optimization in a similar vein as the linear compositions described in Eqs.~\eqref{eq:comb_dae}--\eqref{eq:comb_sf}. This can be seen, by the products between the outer parentheses that surround the  encoding and decoding matrices in Eqs.~\eqref{eq:rest_learn_dec} and~\eqref{eq:rest_learn_enc}. The Hadamard products including the $g'(x)$ inside the first parentheses of Eqs.~\eqref{eq:rest_learn_dec} and~\eqref{eq:rest_learn_enc} can be understood as the search of elements contained in the decoder and encoder matrices that  contribute to the mapping of the mixture to the corresponding output. From the above, it can be said that the compositional strategy applies a layer-wise restriction. This is in contrast to the student strategy, which does not take into account the information in each $\mathbf{W}_{*}$ and $\mathbf{b}_{*}$. The layer-wise restriction forces the couplings matrix to be computed using the parameters of each model, similar to the linear composition that algebraically corresponds to the mapping function. It should be mentioned that regardless of the strategy, i.e., student or compositional, the computed matrices $\mathbf{C}_{*}$ are allowed to retain negative values. That is because the destructive property of the negatives values during the computation of the vector-matrix products are helpful in estimating the target source's magnitude information. The pseudo-algorithm of the NCA for both strategies is given in
Algorithm~\ref{algo:nca}.

{\begin{center} 
\begin{minipage}{.7\linewidth}
\begin{algorithm}[H]
\caption{The Neural Couplings Algorithm}
\begin{algorithmic}[H]
    \REQUIRE{Mixture spectra $\tilde{\mathbf{x}}$, model $\mathcal{M}(\cdot)$, model's parameters $\mathbf{W}_{*}, \mathbf{b}_{*}$, $N \times N$ identity matrix $\mathbf{I}_{N}$, total number of layers in model $L'$, number of iterations $N_{it}$, strategy $S \in \{\text{student, compositional}\}$}, random generator function $rnd$, optimizer/solver $\mathcal{A}(\cdot)$\\
    \STATE{$\mathbf{y}\gets \mathcal{M}(\tilde{\mathbf{x}})$}
    \IF{$S$ is {\em student}}
        \STATE $\mathbf{C}\gets rnd$ 
    \ELSE
        \STATE $\mathbf{C} \gets \mathbf{I}_{N}$
        \FOR{$l':=1$ \TO $L'$}
            \STATE $\mathbf{P}_{l'}\gets rnd$
            \STATE $\mathbf{G}_{l'}\gets g(\mathbf{P}_{l'}(\mathbf{W}_{l'}+\mathbf{b}_{l'})^{T})$
            \STATE $\mathbf{C} \gets (\mathbf{W}_{l'} \odot \mathbf{G}_{l'})\mathbf{C}$
        \ENDFOR
    \ENDIF
    \FOR{$i:=1$ \TO $N_{it}$}
        \STATE $E \gets ||\mathbf{y} - \mathbf{C}\tilde{\mathbf{x}}||$
        \IF{$S$ is {\em student}}
            \STATE $\mathbf{C} \gets \mathbf{C} -  \mathcal{A}(\frac{\partial E}{\partial \mathbf{C}})$
        \ELSE
            \STATE $\mathbf{C} \gets \mathbf{I}_{N}$
            \FOR{$l':=1$ \TO $L'$}
                \STATE $\mathbf{P}_{l'}\gets\mathbf{P}_{l'} -  \mathcal{A}(\frac{\partial E}{\partial \mathbf{P}_{l'}})$
                \STATE $\mathbf{G}_{l'}\gets g(\mathbf{P}_{l'}(\mathbf{W}_{l'}+\mathbf{b}_{l'})^{T})$
                \STATE $\mathbf{C} \gets (\mathbf{W}_{l'} \odot \mathbf{G}_{l'})\mathbf{C}$
            \ENDFOR 
        \ENDIF
        \STATE $\mathbf{C}^{\mathit{o}} \gets \mathbf{C}$
    \ENDFOR \\
    \RETURN $\mathbf{C}^{\mathit{o}}$\\
\end{algorithmic}
\label{algo:nca}
\end{algorithm}
\end{minipage}
\end{center}}
%
%
\section{Experimental Procedure}\label{sec:experiments}
%
%
\subsection{Training \& Assessing the Source Separation Models}
\label{subsec:models_ta}
To optimize the parameters contained in Eqs.~\eqref{eq:dae}--\eqref{eq:sf}, we use the $100$ two-channel multi-tracks available in the MUSDB18 data-set~\cite{musdb18} that were used in the SiSEC 2018 campaign~\cite{sisec18}. The multi-track recordings are sampled at $44100$ Hz. For each multi-track, we use the mixture and singing voice signals in the data-set, and average the two available channels, i.e., monaural mixing.
To construct the data-set $\mathcal{D}$, the STFT analysis is performed for each mixture and corresponding source signal, using a hamming windowing $46$ ms long, a factor of $2$ for zero-padding, a hop-size of $8.7$ ms, and a frequency analysis of $N' = 4096$ from which only the first $N = 2049$ frequency sub-bands are retained (due to the redundancies of the discrete Fourier decomposition). After computing the magnitude of the complex representation, each frequency sub-band is normalized to have a unit variance with respect to the time frames.

We use a single encoding and decoding layer for all models in all approaches. The number of hidden layers for MSS-DAE is set to $L=2$ (MSS-DAE has $L' = 4$ layers in total). The dimensionality through the layers is preserved the same in order to avoid any implicit model regularization~\cite{vincent_den, vincent_08_den, conservative_autoencoders}. The number of layers for the MSS-DAE model was chosen experimentally according to the saturation in minimizing Eq.~(\ref{eq:mse}) during the training process, with respect to the number of hidden layers. All the weight matrices are initialized with samples drawn from a normal distribution and scaled by $\sqrt{\frac{1}{N}}$ as proposed in~\cite{glorot}. The bias terms are initialized to zero. The data-set $\mathcal{D}$ is randomly shuffled, and the training is performed using batches of $128$ time-frames. For gradient-based optimization, the Adam algorithm~\cite{adam} is used with the initial learning rate set to $1e-3$, and decreased by half if no improvement to the loss was observed for two consecutive iterations over all available training data. The exponential decay rates for the first and second-order moments of the Adam algorithm are set to $0.9$ and $0.999$, respectively, following the proposed settings presented in~\cite{adam}. The training is terminated after no improvement is observed over four consecutive iterations throughout the training data. We iterate through the whole model training procedure $50$ times using different random initialization states. This is performed in an attempt to minimize the induced bias of randomly initializing the models' layers, thus more reliably addressing our second research question. All our experiments are carried out using NVIDIAs GeForce GTX Titan X and the PyTorch framework\footnote{\url{https://pytorch.org/}}.

To assess the source separation models, we focus on evaluating the ability of each model to exploit the structure in music spectral representations. To do so, we use the couplings matrix computed with the NCA,  which acts as a data and model-specific filtering operator. According to~\cite{wiener_matrix_guy}, there are two broad classes of filtering operators. The first class is the {\em vector} filtering operator, where the matrix responsible for filtering, i.e., the couplings matrix in our case, contains high values of magnitude on off-diagonal elements. The main benefit of off-diagonal elements is that they allow the exploitation of inter-frequency relationships of the spectral data. In contrast, the second class of filtering operators is denoted as {\em scalar} filters. Scalar filters are characterized by magnitude only on the main diagonal of the couplings matrix. Each element on the main diagonal scales individually the corresponding frequency sub-band. The latter operation is equivalent to the application of a masking strategy~\cite{ps_masks}. 
In practice, source separation models are optimized using  many training examples. Consequently, learning a mapping function with activity on the main diagonal could imply a limited performance in estimating the singing voice spectra. 

Based on the previous argument, we define an objective measure   denoted  the {\em trace-to-off-diagonal-ratio} (TOD-R). The TOD-R is computed as follows:
\begin{equation}
\text{TOD-R}(\mathbf{C}_{*}) = \sqrt{N} \frac{\text{tr}(|\mathbf{C}_{*}|)}{|| \mathbf{C}_{*} \odot (\mathbf{J}_{N} - \mathbf{I}_{N})||},
\label{eq:todr}
\end{equation}
where $\text{tr}(\cdot)$ is the trace function that adds up all the elements on the main diagonal of the matrix, $\mathbf{I}_{N}$ is the $N \times N$ identity matrix, and $\mathbf{J}_{N}$ is the $N \times N$ with all elements equal to one. The scaling by $\sqrt{N}$ is performed in order to compensate for the initialization scaling described before, and due to the expected high values of the denominator in Eq.~\eqref{eq:todr} (since the norm is taken using far more matrix elements than the trace in the numerator). The element-wise absolute $|\cdot|$ is computed prior to the computation of the trace to avoid biasing the ratio due to the norm (sum of absolute values) in the denominator of Eq.~\eqref{eq:todr}. Small values of TOD-R indicate that the off-diagonal elements of the couplings matrix retain higher magnitude values than the elements on the main diagonal and vice versa. High off-diagonal activity suggests that the mapping function of the model exploits more inter-frequency relationships rather than estimating values that scale the mixture spectra as in scalar-based filtering and frequency masking~\cite{wiener_matrix_guy}.

\subsection{Computing \& Assessing the Neural Couplings}
\label{subsec:nca_model_ta}
For computing the neural couplings, i.e., solving the optimization problem in Eq.(\ref{eq:argmin_c}), we use the test subset from the MUSDB18 data-set~\cite{musdb18}. The test subset comprises $50$ additional two-channel, multi-tracks sampled at $44100$ Hz. For computing the STFT we employed the same parameters as during the construction of the training data-set $\mathcal{D}$ reported in Sec.~\ref{subsec:models_ta}. As it is more informative to compute mapping functions that process multiple time-frame vectors rather than a single instance, we use a batch of adjacent magnitude vectors drawn from the test subset. The batch size is set to $T=350$ ($\sim$ 3.1 seconds long), and the hyper-parameter $T$ is experimentally chosen based on the trade-off between the number of vectors and computational resources. With this, we can simply reformulate the NCA using matrix notation instead of vector notation: $\tilde{\mathbf{X}} \in \mathbb{R}_{\geq 0}^{N \times T}$ instead of $\tilde{\mathbf{x}} \in \mathbb{R}_{\geq 0}^{N}$ and ${\mathbf{Y}} \in \mathbb{R}_{\geq 0}^{N \times T}$ instead of ${\mathbf{y}} \in \mathbb{R}_{\geq 0}^{N}$. For the solver denoted as $\mathcal{A}(\cdot)$ in Algorithm~\ref{algo:nca}, we use the Adam algorithm with a learning rate equal to $4e^{-4}$ and the same exponential decay rates as in Sec.~\ref{subsec:models_ta}. The total number of iterations is set to $600$, and the random function ($rnd$ in Algorithm 1) refers to drawing samples from a normal distribution scaled by $\sqrt{\frac{1}{N}}$ as proposed in~\cite{glorot}. The above mentioned hyper-parameters were chosen experimentally.

In comparison to the optimization of the music source separation models presented in Section~\ref{subsec:models_ta}, the couplings matrix is computed for each model and for each batch of adjacent magnitude vectors drawn from the test subset of MUSDB18 data-set~\cite{musdb18}.
The pre-trained parameters of each one of the three models are randomly drawn from one of the $50$ training instances. In our initial experiments it was observed that the silent segments in the used multi-tracks led to randomly structured and sparse, i.e., low $\ell_{1}$ norm values, row-vectors of the computed couplings matrix $\mathbf{C}^{\mathit{o}}$. Although the observed convergence of the NCA was satisfactory for the silent segments, it significantly biased the TOD-R measure in an unpredictable manner. Therefore, from each multi-track we select the $30$ seconds that all music sources available in the data-set are active. The selection of the active waveform regions, is based on the generator\footnote{Available at: https://github.com/sigsep/sigsep-mus-cutlist-generator} used in the music source separation evaluation campaign~\cite{sisec18}.

As we have not provided any theoretical guarantees regarding the convergence of the NCA to an optimal solution, we conduct a quantitative analysis to assess the ability of the NCA to accurately \emph{approximate the models' outputs}. We use the signal to noise ratio (SNR) expressed in dB as an intuitive measure of the quality of the NCA approximations. For the DAE and MSS-DAE models, the SNR is computed using the singing voice magnitude spectra estimated by the NCA, and the singing voice magnitude spectra estimated by the corresponding models. For the SF model, the predicted mask using the NCA is first applied to the input mixture, and the outcome is used to compute the SNR. This is done because the SF model does not employ any pre-computed mask during its optimization that could be used for evaluating the NCA. For baseline comparison, we use the linear composition and identity function as proxies to the couplings matrix, i.e., $\mathbf{C}$ is computed using Eqs.~\eqref{eq:comb_dae}--\eqref{eq:comb_sf} and $\mathbf{C} = \mathbf{I}_{N}$, respectively. Since the SNR is computed using the approximations of the NCA and the models' outputs, different SNR values across the source separation models are expected.

In addition to the above, we also compute the SNR using the NCA approximation and the true source singing voice spectra, and compare it with the SNR using the models' outputs and the true source singing voice spectra. This is done in order to intuitively quantify the loss of information induced by the NCA. The previously mentioned analysis is performed for each strategy, model, and segment in each multi-track. It is important to note that the goal here is to understand the power of the NCA to deliver an accurate approximation of the model's output. With this in mind, we focus our analysis on the SNR and the TOD-R, and leave an analysis of the implications of model choice on separation performance for future endeavours.
%
%
\section{Results \& Discussion}
\label{sec:results}
\begin{figure}[t]
	\centering
	\includegraphics[width=0.75\columnwidth, keepaspectratio, trim={1cm 1cm 2cm 1cm}]{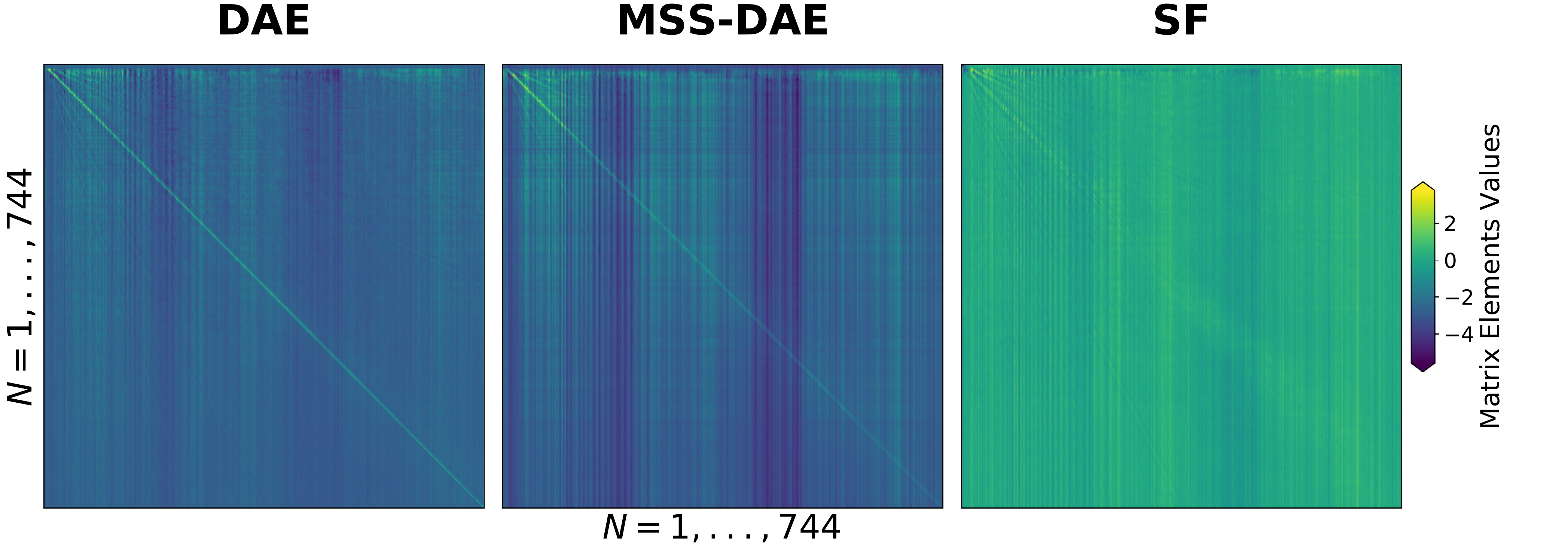}
	\caption{The linear composition of the models' encoding and decoding functions. The compositions are computed using the Eqs.~\eqref{eq:comb_dae}--\eqref{eq:comb_sf}, and averaged across the $50$ experimental iterations.
	The first $744$ frequency sub-bands ($\sim8$kHz) are displayed for clarity.
	{\em Left Column}: Composition for the denoising auto-encoder model~\cite{vincent_08_den} (DAE). {\em Middle Column}: Composition for the four layer extension of the DAE model, adapted to music source separation (MSS-DAE)~\cite{uhl15,nug16}. {\em Right Column}: Composition for the skip connections for filtering the input mixture (SF)~\cite{mim17_mlsp, wening14, jannson17}.}
    \label{fig:neural_comb}
\end{figure}

To address our research question ``\emph{\textbf{RQ2: Do DAEs that are commonly employed in music source separation learn trivial solutions for the given problem}}?'', we compute the linear composition functions for the three models (DAE, MSS-DAE, and SF) using  Eqs.~\eqref{eq:comb_dae}--\eqref{eq:comb_sf}. The computation of the linear compositions follows the research findings presented in~\cite{conservative_autoencoders} that underlines the tendency of encoding and decoding functions to become symmetric, that practically leads to learning scalar filtering operators. The average result across the $50$ experimental iterations from the composition functions is illustrated in~Fig.\ref{fig:neural_comb}. 
By observing Fig.~\ref{fig:neural_comb} it is evident that the two models that map directly the mixture magnitude spectra to the singing voice spectra (DAE and MSS-DAE) have a prominent main diagonal structure. This means that through training, the corresponding encoding and decoding functions tend to become symmetric in order to provide a solution in the MSE sense. Therefore, it is plausible that the DAE and MSS-DAE models learned trivial solutions to the problem of singing voice separation, restricting the overall separation performance. On the other hand, employing the skip-connections as performed in the SF model, it can be observed that the activity has been repelled from the main diagonal (see the right column of Fig.~\ref{fig:neural_comb}). This can be explained by recalling that a solution for estimating the singing voice spectra is the scaling of the individual frequency sub-bands of the mixture, which in turn is expressed by a diagonal matrix. That statement provides a simple explanation on why skip-connections~\cite{joao_skip} and end-to-end learning~\cite{dieleman14}, where the time-domain signals are used instead of spectrograms, are emerging directions in music source separation.

\begin{table}
\centering
\caption{Assessing the mapping functions. The TOD-R metric (Eq.~\eqref{eq:todr}) for each strategy and model.}

\begin{tabular}{c|ccc}
 &  & \textbf{Model} &     \\ \hline
\begin{tabular}[c]{@{}c@{}} \textbf{Strategy}
\end{tabular}     & DAE   & MSS-DAE & SF \\ \hline
\begin{tabular}[c]{@{}c@{}}Student\end{tabular} & $0.03\;(\pm0.00)$ & $0.03\;(\pm 0.00)$ & $\mathbf{0.02\;(\pm 0.00)}$  \\
\begin{tabular}[c]{@{}c@{}}Compositional\end{tabular} & $0.36\;(\pm 0.13)$ & $0.14\;(\pm 0.04)$ & $\mathbf{0.03\;(\pm 0.01)}$  
\end{tabular}
\label{table:tod-res}
\end{table}

Aiming to address the other research question ``\emph{\textbf{RQ1: Why is masking  important in approaches based on the DAE model}}?'', we turn our focus to the computed mapping functions using the NCA, and the model assessment via the TOD-R metric. Table~\ref{table:tod-res} summarizes the TOD-R results for each model, both for the student and the compositional strategies. The average TOD-R across all the segments of the test-subset is reported, and in the parentheses the standard deviation of the corresponding measurements is provided. Bold faced numbers, indicate the smallest obtained value for the TOD-R implying that the model has exploited a richer inter-frequency structure.

The results of Table~\ref{table:tod-res} show that the SF model provides the smallest TOD-R value for both strategies. More specifically, the TOD-R for the compositional strategy and the SF model is 12 times smaller than  the DAE model, which has equal number of encoding and decoding layers. In comparison to the MSS-DAE model that comprises two additional hidden layers, the TOD-R value of the SF model is decreased by approximately $4.6$ times. We could conclude that skip connections can be seen as a simple method to repel source separation approaches from learning solutions that concentrate most of the activity on the main diagonal of their corresponding mapping function(s).

Additionally, Table~\ref{table:snr} presents the results from the evaluation of the approximation performance of the NCA strategies by means of the SNR. The corresponding values are reported in dB. The average approximation performance of the NCA for the DAE and SF models and both strategies, outperforms the linear composition by approximately $6$dB, and by a very large margin, greater than $100$ dB, in the case of the MSS-DAE model. Experimental observations suggest that the large margin is due to the exceeding norm of the spectra approximated by the linear composition. The exceeding norm of the spectra is itself due to the high norm of the row-vectors of $\mathbf{C}$, computed using the linear composition, that are used to approximated the MSS-DAE's output(s). This in turn, shows that by increasing the number of layers in non-linear source separation models, the linear composition is a poor proxy of the models' mapping functions. As for the identity function, i.e., using the mixture instead of the models' output spectral estimates, it is outperformed by the NCA on average, across strategies and models, by $\sim4$dB. The student strategy outperforms the compositional strategy on average across the source separation models by $0.7$ dB. This is due to the fact that the student strategy employs gradient updates that are related only to the reconstruction error minimization.

\begin{figure}
\vspace{-1.cm}
\begin{minipage}[t]{0.49\columnwidth}
\includegraphics[width=1.\columnwidth]{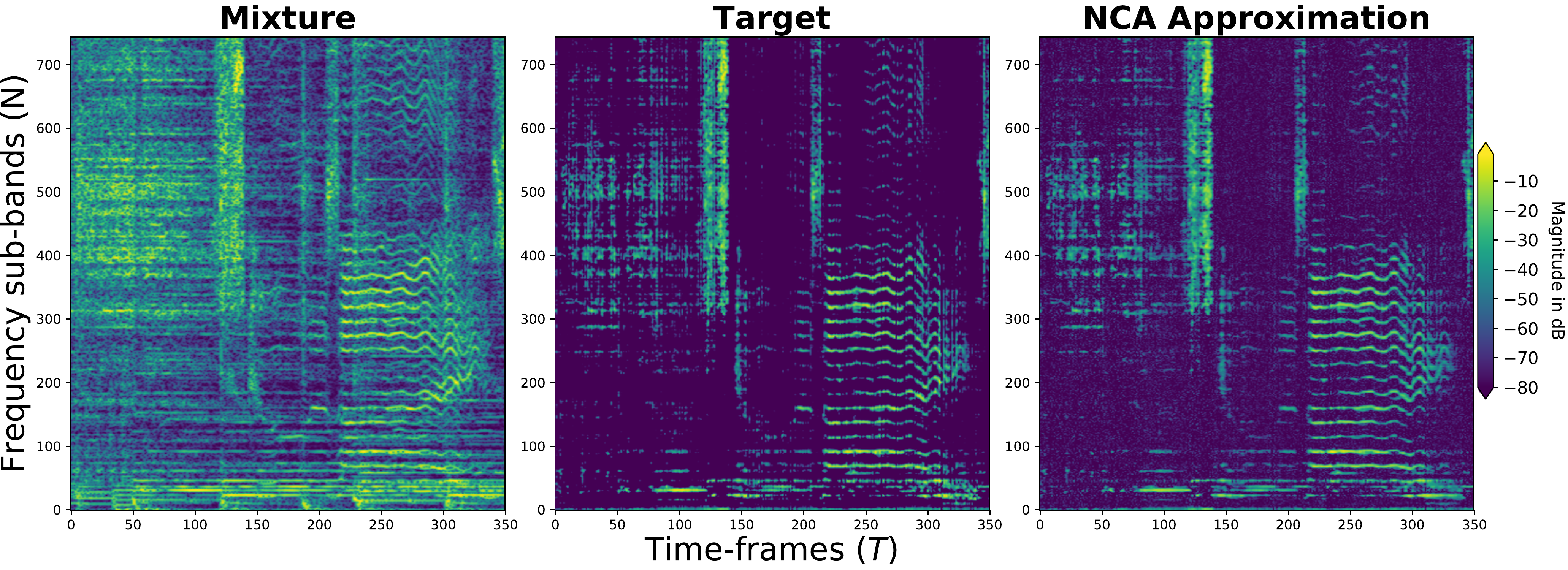}
\includegraphics[width=0.85\columnwidth, height=3.45cm]{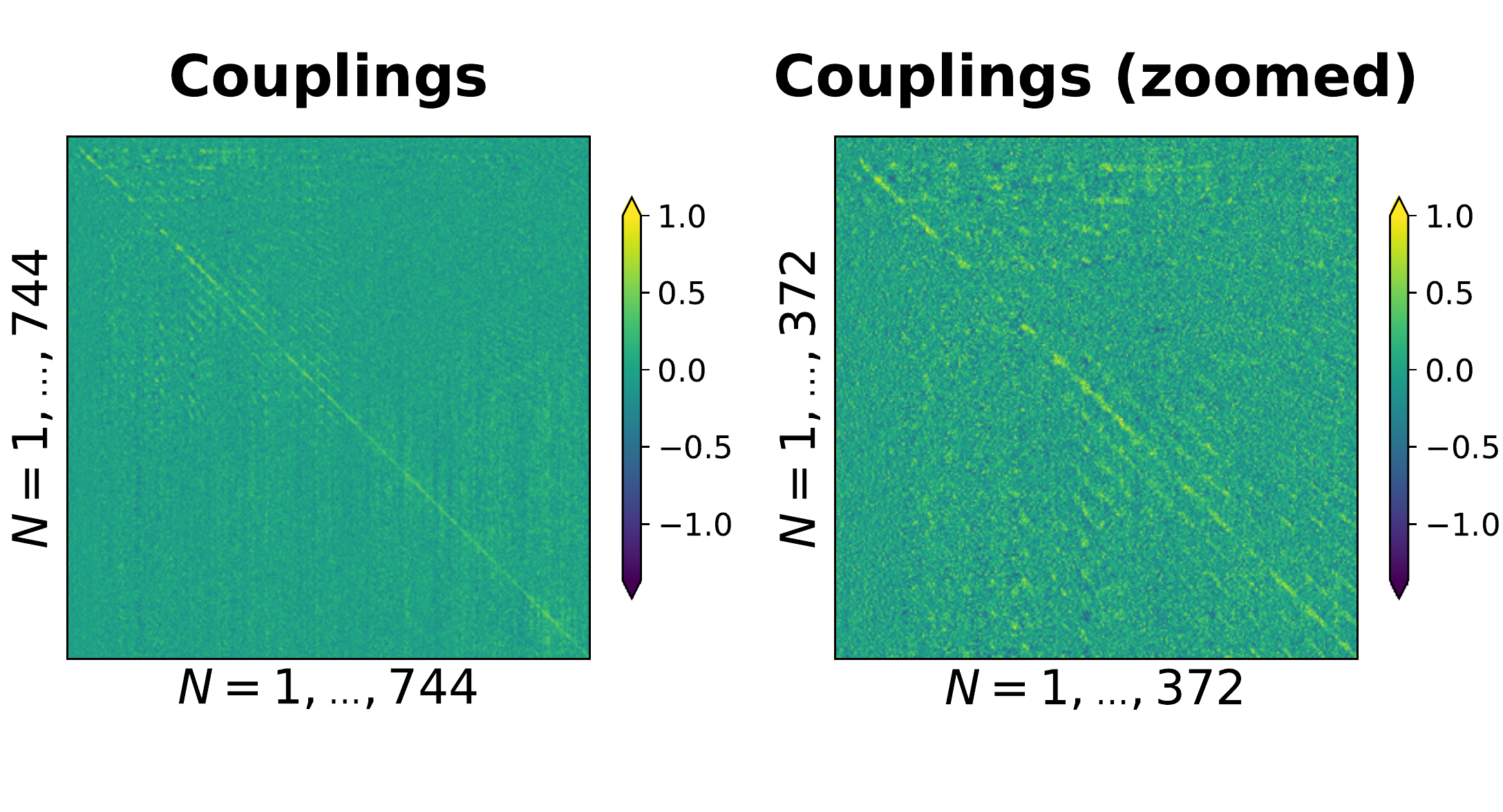}
\centering
{\tiny \subcaption{Student: DAE}}
\end{minipage}
\begin{minipage}[t]{0.49\columnwidth}
\centering
\includegraphics[width=1.\columnwidth]{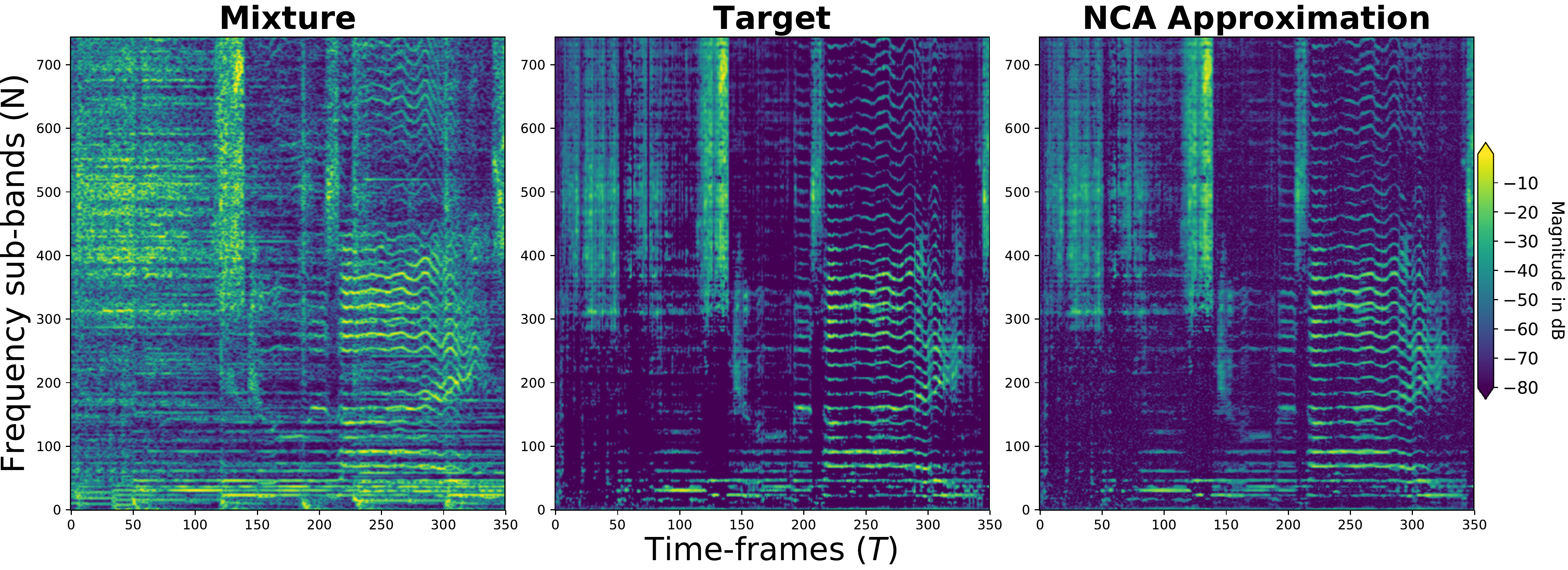}
\includegraphics[width=0.85\columnwidth, height=3.45cm]{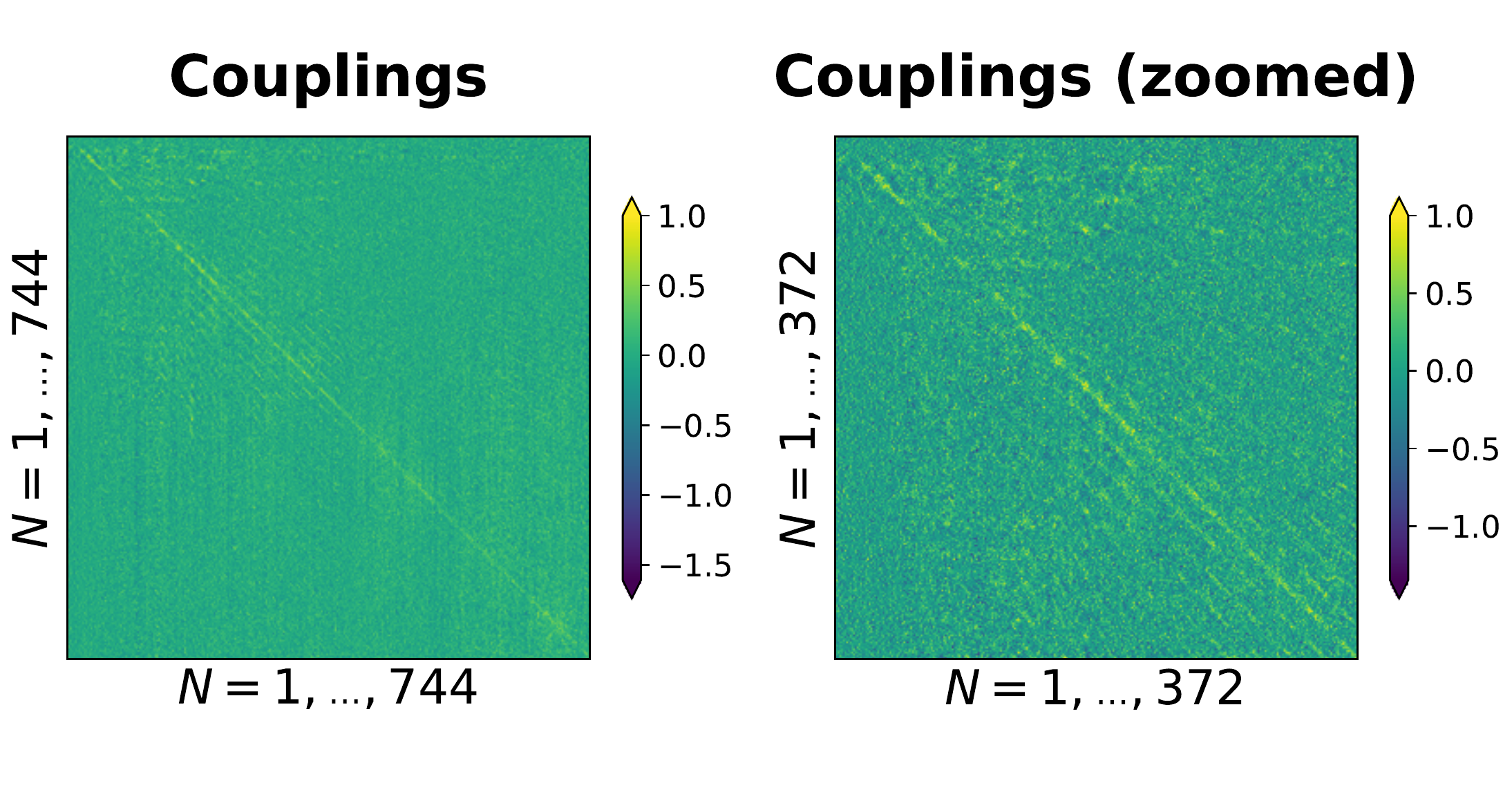}
{\tiny \subcaption{Student: MSS-DAE}}
\end{minipage}
\begin{minipage}[t]{0.49\columnwidth}
\centering
\includegraphics[width=1.\columnwidth]{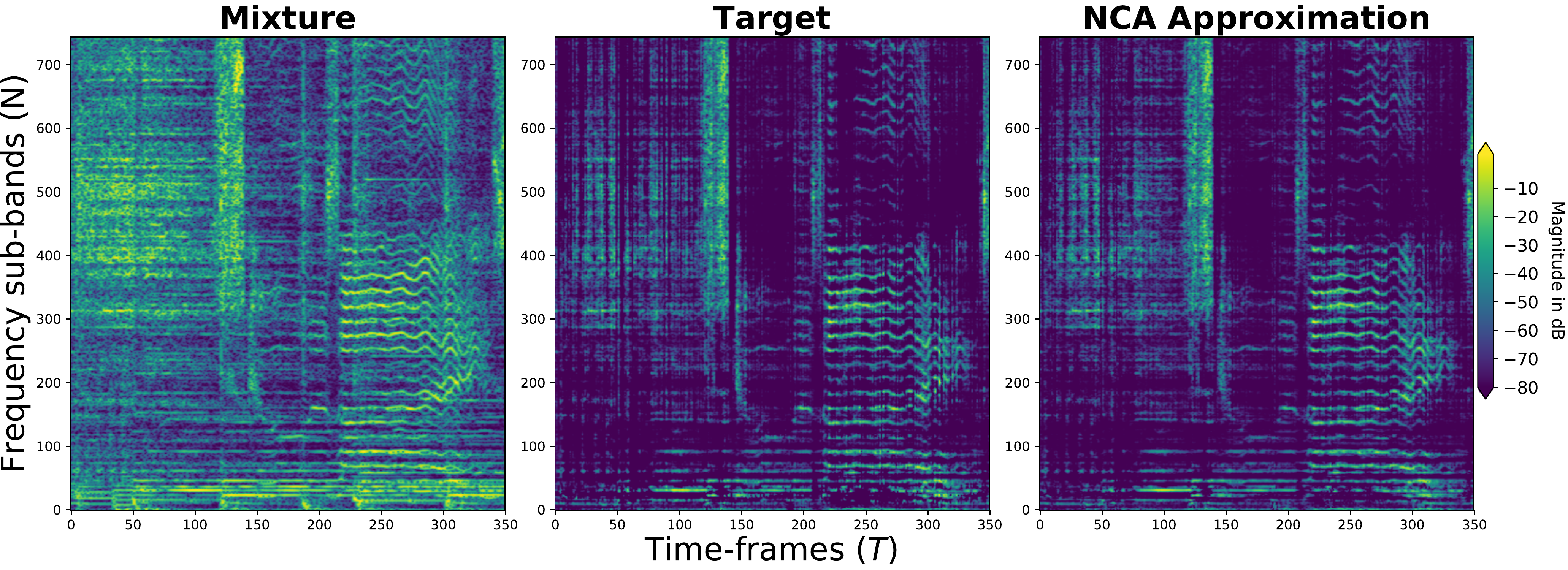}
\includegraphics[width=0.85\columnwidth, height=3.45cm]{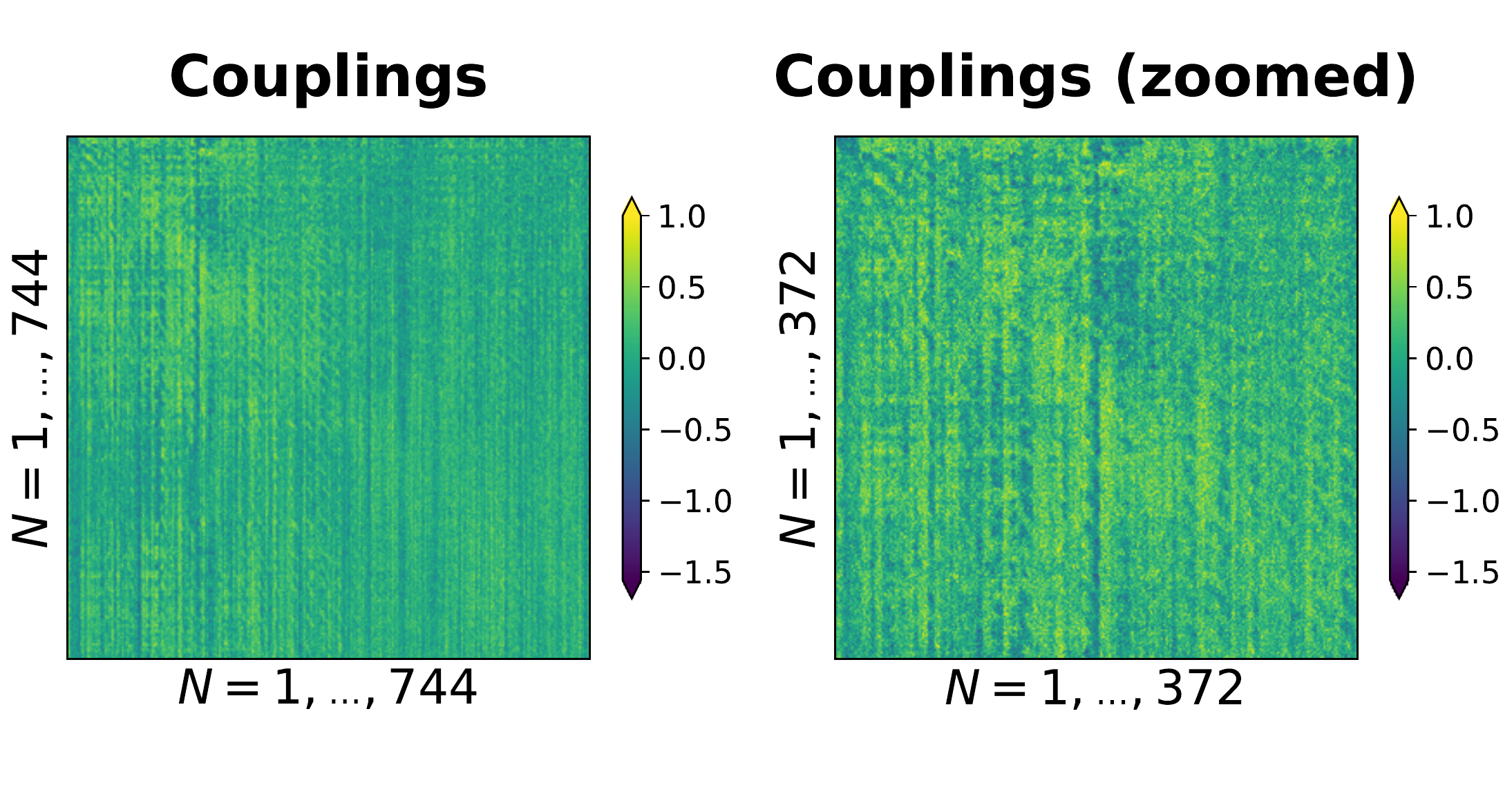}
{\tiny \subcaption{Student: SF}}
\end{minipage}
\begin{minipage}[t]{0.49\columnwidth}
\centering
\includegraphics[width=1.\columnwidth]{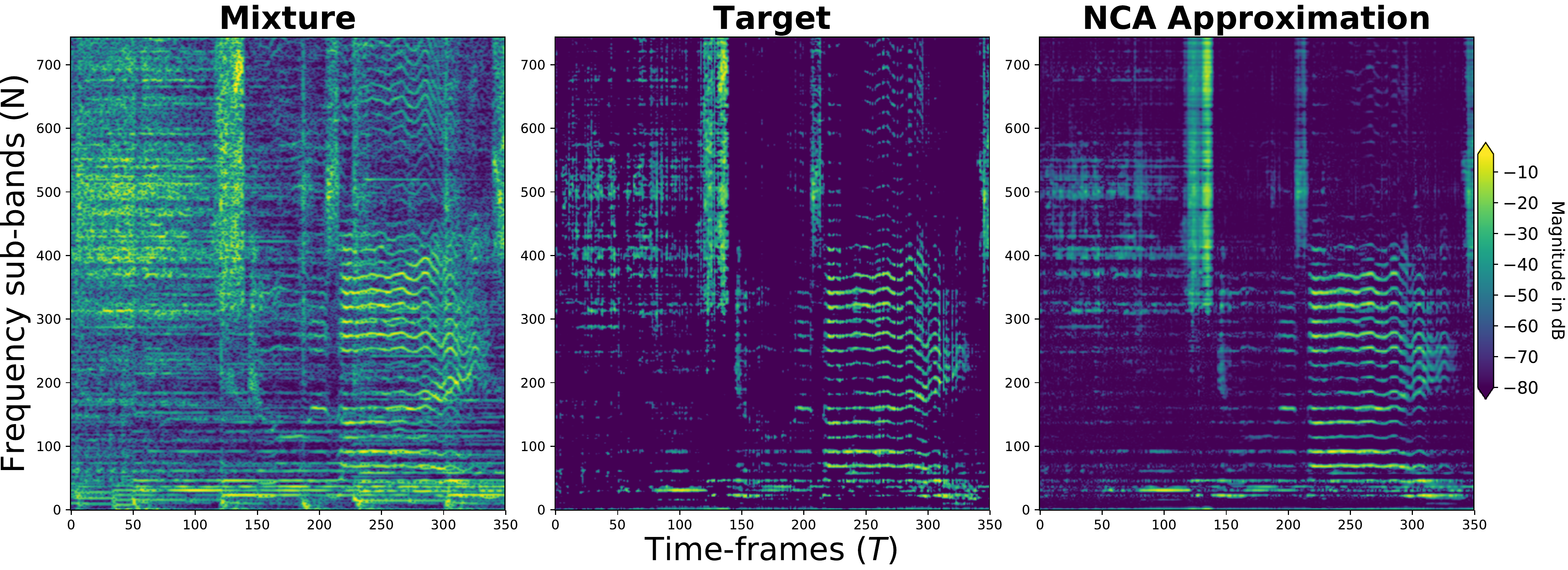}
\includegraphics[width=0.85\columnwidth, height=3.45cm]{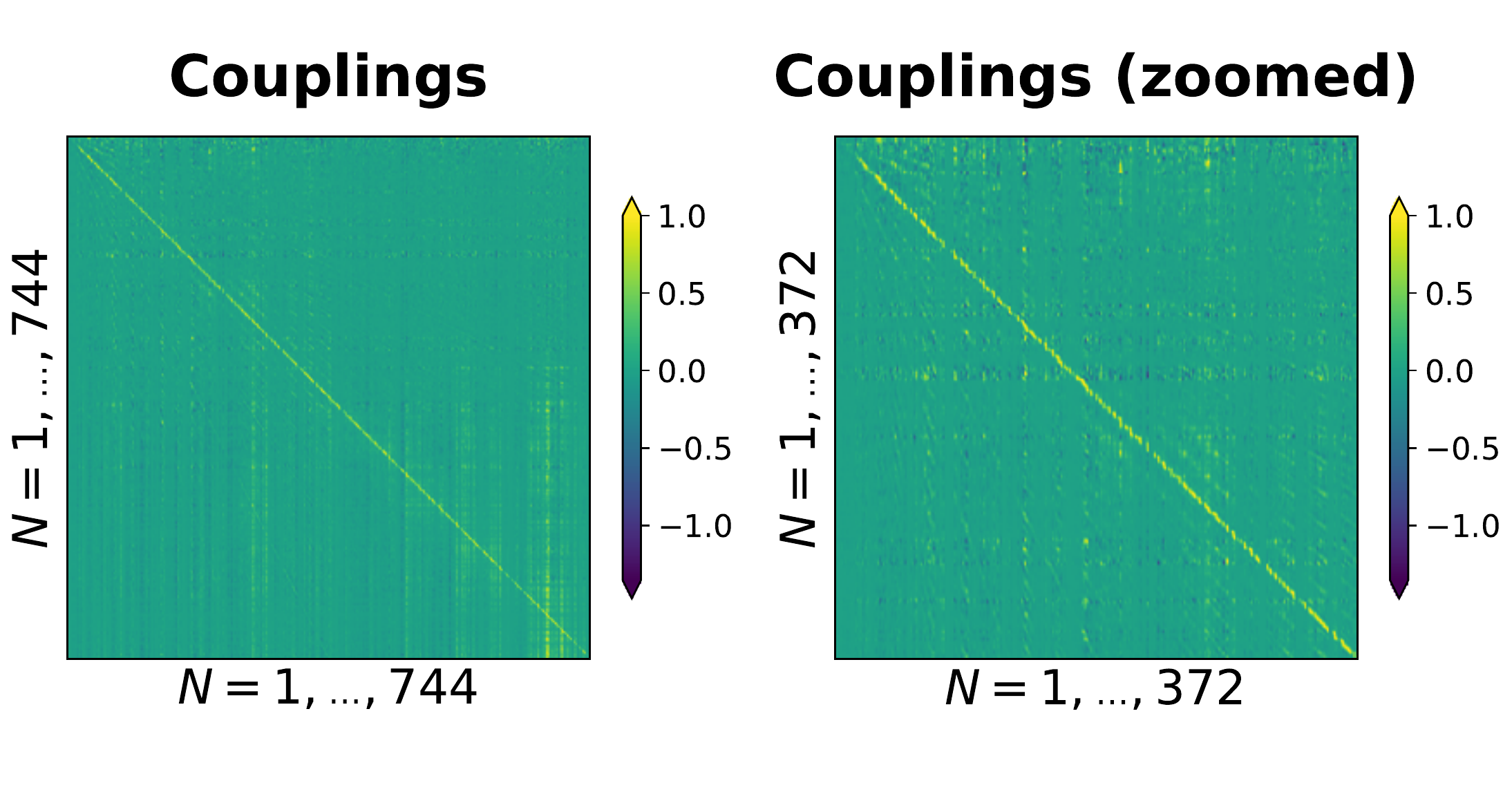}
\centering
{\tiny \subcaption{Compositional: DAE}}
\end{minipage}
\begin{minipage}[t]{0.49\columnwidth}
\centering
\includegraphics[width=1.\columnwidth]{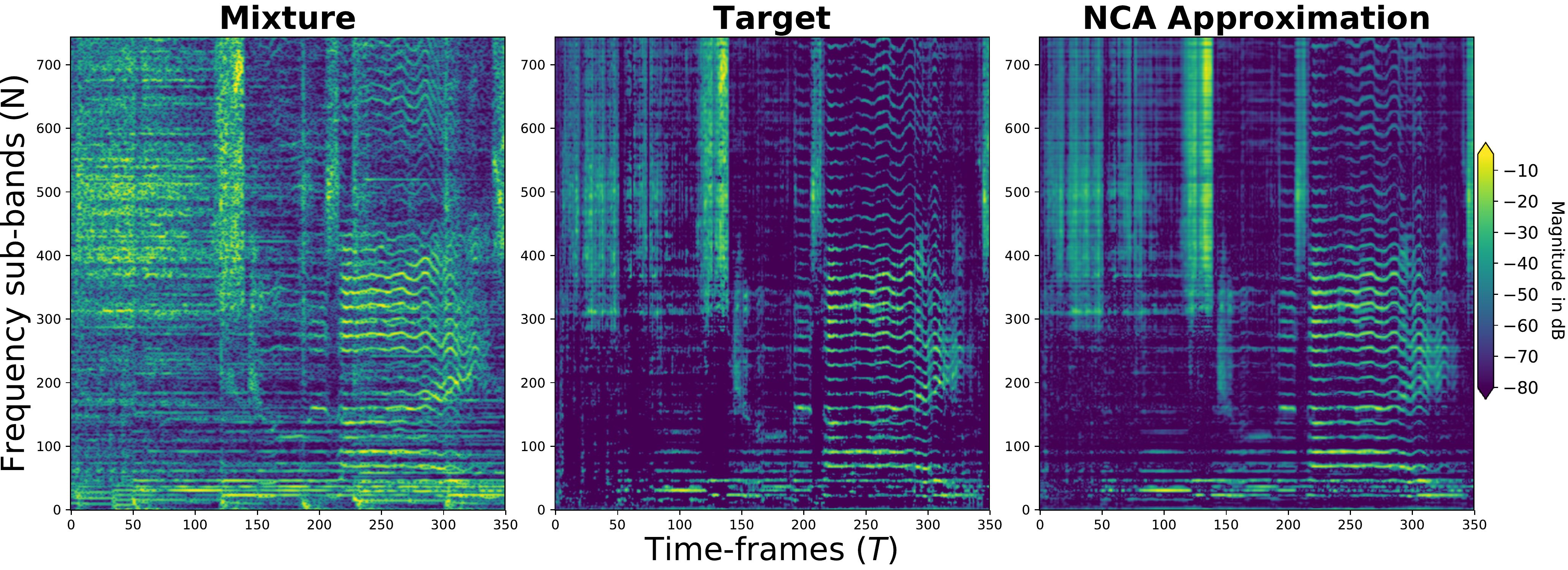}
\includegraphics[width=0.85\columnwidth, height=3.45cm]{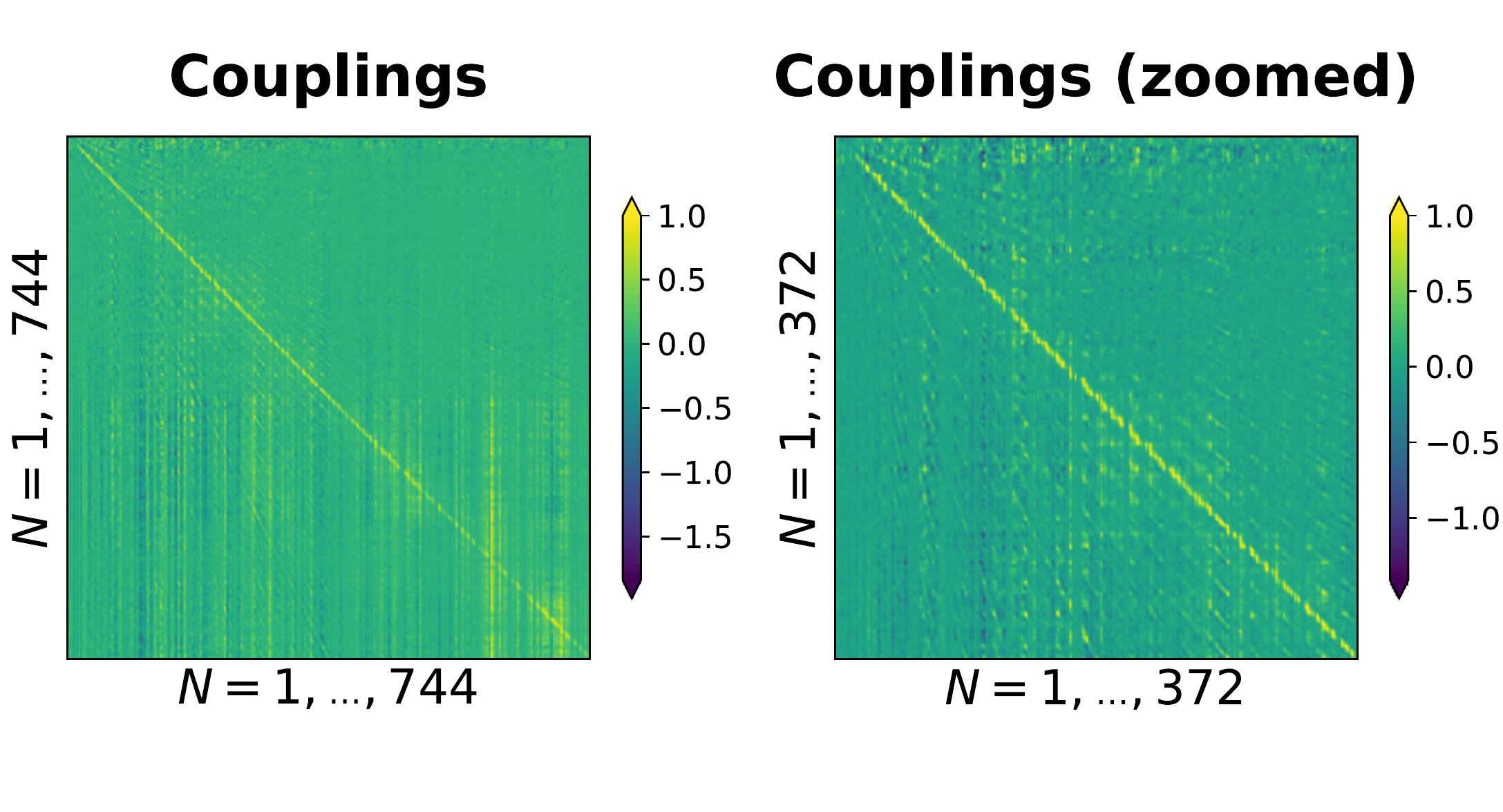}
{\tiny \subcaption{Compositional: MSS-DAE}}
\end{minipage}
\begin{minipage}[t]{0.49\columnwidth}
\centering
\includegraphics[width=1.\columnwidth]{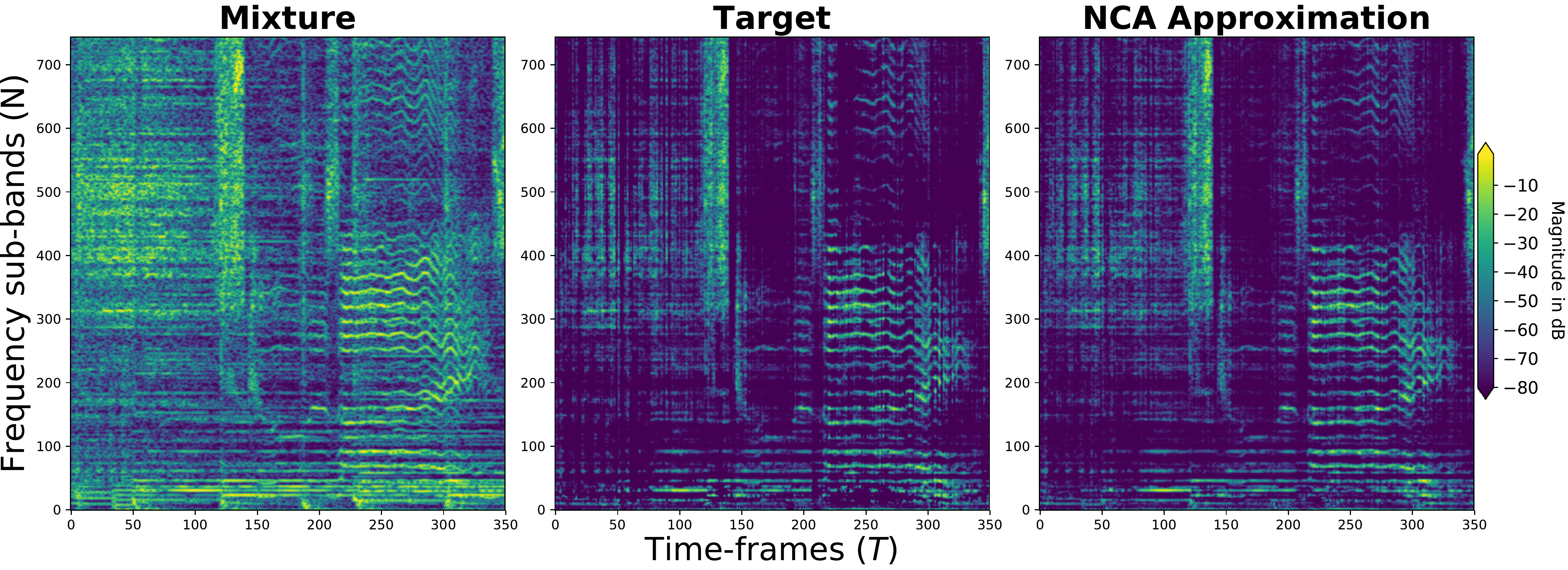}
\includegraphics[width=0.85\columnwidth, height=3.45cm]{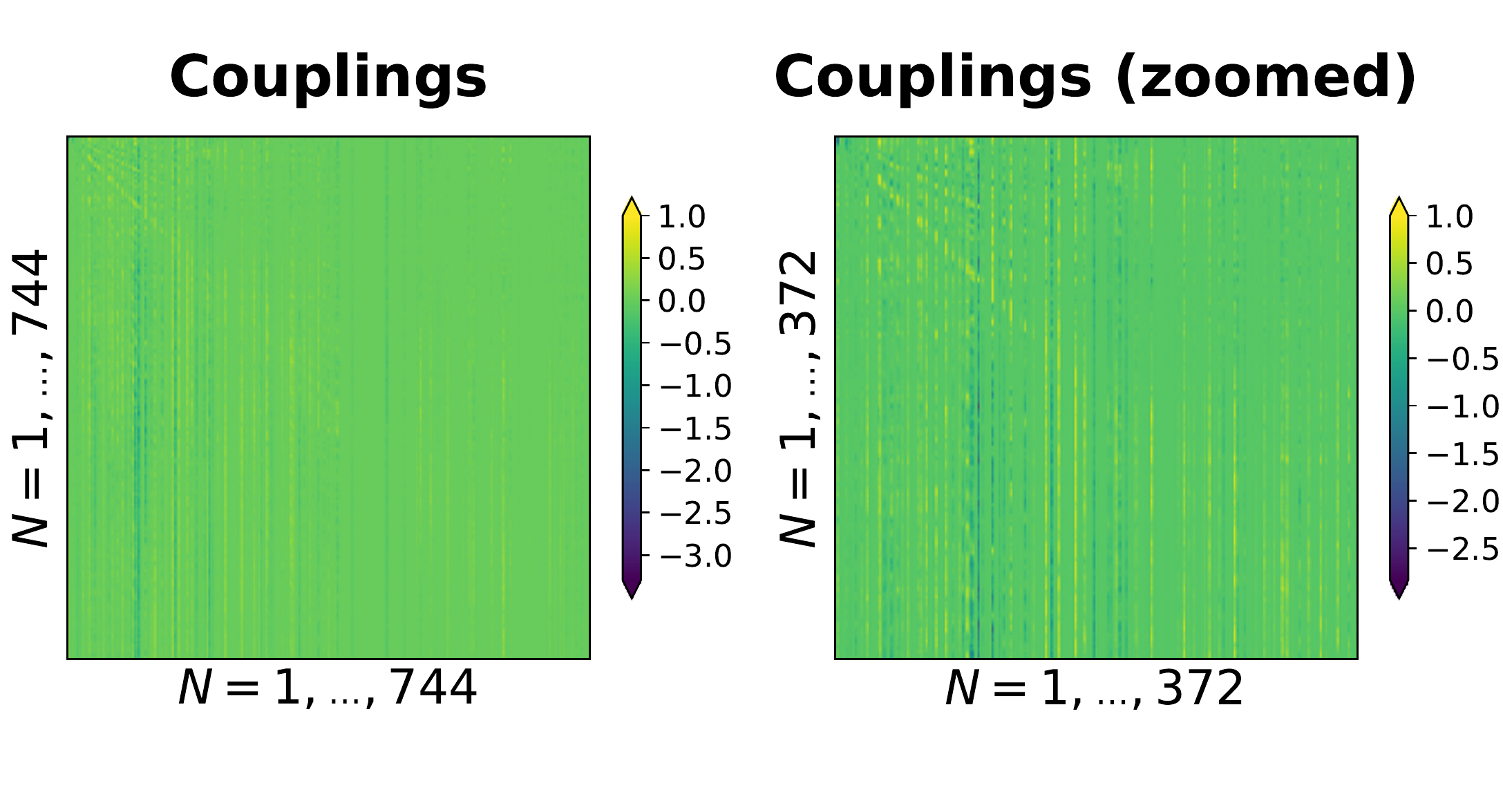}
{\tiny \subcaption{Compositional: SF}}
\end{minipage}
\vspace{-0.13cm}
\caption{The outcome of the NCA for the DAE, MSS-DAE, and SF models using a $\sim3$ seconds excerpt from the file \textit{Al James - Schoolboy Fascination}  in the test sub-set of MUSDB18. {\em First Row (a)--(c):} The couplings matrices approximating the mapping functions and the corresponding spectral estimates using the {\em student} strategy. {\em Second Row: (d)--(f):} The couplings matrices approximating the mapping functions and the corresponding spectral estimates using the {\em compositional} strategy. A row-wise maximum value normalization, and zooming of the couplings matrices is presented for clarity.}
\label{fig:couplings_figs}
\end{figure}

\begin{table}
\caption{Assessing the approximation performance of the mappings computed by the NCA, compared to the models' outputs. The mean and standard deviation of the SNR, expressed in dB, are reported. For comparison, the linear composition and identity function are used. Bold faced values denote the best approximation performance.}
\centering
\begin{tabular}{c|ccc}
 &  & \textbf{Model} &     \\ \hline
\begin{tabular}[c]{@{}c@{}} \textbf{Strategy}
\end{tabular}     & DAE   & MSS-DAE & SF \\ \hline
\begin{tabular}[c]{@{}c@{}}Student\end{tabular}       & $\mathbf{6.51(\pm 1.79)}$     & $9.11(\pm 1.09)$ & $\mathbf{6.53(\pm 2.31)}$    \\
\begin{tabular}[c]{@{}c@{}}Compositional\end{tabular} & $4.78(\pm 2.69)$     & $\mathbf{9.25(\pm 1.13)}$ & $5.98(\pm 2.24)$  \\
\end{tabular}
\begin{tabular}{c|ccc}
\hline
\begin{tabular}[c]{@{}c@{}} \textbf{Baseline}
\end{tabular}     & DAE   & MSS-DAE & SF \\ \hline
\begin{tabular}[c]{@{}c@{}}Linear Comp. \end{tabular} & $0.01(\pm 0.01)$ & $-174.9(\pm 18.4)$ & $0(\pm 0)$ \\
\begin{tabular}[c]{@{}c@{}}Identity Funct.\hspace{-0.05cm} \end{tabular} &$2.38(\pm 0.81)$ & $3.05(\pm 0.66)$ & $2.51(\pm 1.03)$ 
\end{tabular}\\
\label{table:snr}
\end{table}

\begin{table}
\centering
\caption{Assessing the approximation performance of the mappings computed by the NCA, compared to the true singing voice spectra. The mean and standard deviation of the SNR, expressed in dB, are reported. For comparison, the models' output is also reported.}
\begin{tabular}{c|ccc}
 &  & \textbf{Model} &     \\ \hline
\begin{tabular}[c]{@{}c@{}} \textbf{Strategy}
\end{tabular}     & DAE   & MSS-DAE & SF \\ \hline
\begin{tabular}[c]{@{}c@{}}Student\end{tabular}       & $-4.02(\pm 2.43)$     & $-1.94(\pm 2.03)$ & $-4.88(\pm 3.06)$    \\
\begin{tabular}[c]{@{}c@{}}Compositional\end{tabular} & $-5.50(\pm 3.41)$     & $-2.36(\pm 2.29)$ & $-4.15(\pm 3.41)$  \\ \hline
\begin{tabular}[c]{@{}c@{}}\textbf{Model Output}\end{tabular} & $-5.05(\pm 3.05)$     & $-2.17(\pm 2.43)$ & $-4.74(\pm 3.15)$  \\
\end{tabular}
\label{table:snr2}
\end{table}

To further understand the approximation performance of the NCA, we also calculate the SNR values between the NCA approximation and the true singing voice spectra (see Table~\ref{table:snr2}). Results suggest that the NCA approximations are marginally better than those obtained directly with the models' outputs. More specifically and on average across strategies and models, a small improvement of $\sim0.17$ dB is observed.

Focusing on the structure of the mapping functions that are computed using the NCA via the student strategy, illustrated in the first row of Fig.~\ref{fig:couplings_figs}, it can be seen that the couplings matrices, serving as the corresponding mapping functions, are nearly identical between the DAE and MSS-DAE models, and marginally different from the SF model. The marginal differences can be explained by the fact that the couplings matrix for the SF model was optimized according to Eq.~(\ref{eq:argmin_c}) using the output masks of the SF model as target function(s) $\mathbf{Y}$, as opposed to the DAE and MSS-DAE models that use the corresponding singing voice spectral estimates as target function(s).

The above tendency is also demonstrated in Table~\ref{table:tod-res}, where the statistics of the TOD-R metric for the student strategy are exactly the same for the DAE and the MSS-DAE models. This shows that the mappings are nearly identical between the models. This can be explained by recalling the Eqs.~\eqref{eq:argmin_c}--\eqref{eq:dist_pd} that depict the approximation of the mapping functions by observing only input-output and model-dependent relationships of spectral representations, and by neglecting the parameters of the model. The disuse of the model's parameters during the optimization of the NCA is a convex problem that is experimentally shown to lead to similar solution during the experimental realization. However, according to~\cite{wang16_datajacobian} the disuse of the model parameters leads to system solutions that do not characterize the functionality of the non-linear model. Consequently, we turn our focus on the compositional strategy, that includes the knowledge of the parameters of each model. 

In the second row of Table~\ref{table:tod-res} and in Figs.~\ref{fig:couplings_figs}d--\ref{fig:couplings_figs}e, a pattern that is evident among the mapping functions of the DAE and MSS-DAE models is the high diagonal activity when compared to the SF model that employs the skip connections. Specifically, the mapping function of the SF model has pushed most of its activity away from the main diagonal. For small $N$, i.e., in the first matrix rows of the zoomed mapping function of~Fig.~\ref{fig:couplings_figs}f, it can be seen that elementary spectral structures are formed. Those spectral structures are related to the target spectra of~Fig.\ref{fig:couplings_figs}f as both the spectral and the mapping function illustrations concentrate magnitude information in same frequency sub-band regions. According to the graphical model and its expected conditional, i.e., Fig.~\ref{fig:pgm}c and $q(x|\tilde{x})q(\tilde{x})$, we can underline that the mapping function of the SF model captures the relevant structure of the target source that has been observed in the mixture such that a time-frequency mask can be applied to suppress the interfering sources. What is not evident in our results, is the SF model's capacity in learning spectral structures from the training data.

To this end, Fig.~\ref{fig:couplings_figs} also shows that the additional layers that the MSS-DAE model employs are essentially used to model additional inter-frequency relationships, compared to the DAE model that comprises only two layers and concentrates most of its activity on the main diagonal. The MSS-DAE model not only pushes its activity to off-diagonal elements but also forms a structured matrix, mostly seen in the zoomed couplings matrix of Fig.~\ref{fig:couplings_figs}e, that is {\em roughly} similar to a circulant matrix with sparse entries. Those types of matrices are commonly used in digital signal processing for convolutional operators. This observation somewhat justifies the advantage of incorporating additional computational layers into a source separation model as proposed in~\cite{nug16, uhl15}, and serves as an explanation on why convolutional layers are attractive choices in source separation models~\cite{takahashi17}. On the other hand, the DAE model has a simpler structure similar to a band matrix, with the minor off-diagonal activity denoting the spectral relationships, i.e., quasi-harmonic structures of the singing voice.

The code for reproducing the above results and computing the gradients can be found under: \url{https://zenodo.org/record/2629650} and \url{https://github.com/Js-Mim/nca_mss}.
%
%
\section{Conclusions \& Future Research}
\label{sec:conclusions}
In this study, we formalized two research questions
related to  the most commonly used neural network model in music source separation, i.e., the denoising autoencoder presented in~\cite{vincent_08_den}: \emph{\textbf{RQ1}})``\emph{\textbf{Why is masking important in approaches based on the DAE model}}?'', and \emph{\textbf{RQ2}}) ``\emph{\textbf{Do DAEs that are commonly employed in music source separation learn trivial solutions for the given problem}}?''. To answer those questions we focused on the examination of the mapping functions of the corresponding models applied to the particular problem of singing voice separation. For computing the mapping functions, we proposed an experimentally derived algorithm and investigated two strategies to that aim. The first one, denoted as the {\em student strategy}, is based on the neural network distillation concept presented in~\cite{hinton_distill}. It was observed that the student strategy leads to ambiguous results regarding the approximation of the mapping functions as it does not account for the  model's parameters. As an alternative, the {\em compositional strategy} was proposed for taking into account the model's already optimized parameters for the problem of singing voice separation, similarly to the algebraic derivation of the mapping function.

Using the compositional strategy and the computed mapping functions we investigated the denoising autoencoder (DAE) model, its multi-layered extension employed in relevant tasks (MSS-DAE)~\cite{nug16, uhl15, uhl17}, and the denoising autoencoder with skip-connections (SF) that are used to mask the mixture spectra similarly to a time-frequency filtering operation~\cite{mim17_mlsp, wening14, jannson17}. By examining the overall structure of the mapping functions, we conclude that the source separation models learn data-driven filtering functions when they are optimized for singing voice separation. The DAE model learns trivial solutions because the corresponding encoding and decoding functions become symmetric during the training procedure. Consequently, the filtering functions learned by the DAE  act as scalar filters in the frequency domain potentially limiting the overall source estimation performance. Furthermore, employing skip-connections as in the SF model  can be seen as a simple method to enforce DAEs to learn richer inter-frequency dependencies compared to the DAE. This can justify the empirically observed performance boost over DAEs in previous works like~\cite{wening14, mim17, jannson17, drossos18}. Finally, the additional computational layers employed in the MSS-DAE model, seem to promote the learning of filter kernels with a sparse and circulant structure roughly similar to convolutional operations. However, those kernels share also similarities with scalar filters, as in the case of the DAE, which reduce the overall filtering performance and support the experimental results in~\cite{uhl15} that promote the usage of masking as a post-processing step for estimating the target source(s).

Although this study does not fully reflect the current trends in deep learning-based music source separation, it serves as a first step towards understanding what non-linear models learn from data when they are optimized to separate the singing voice. Furthermore, we investigated and provided the graphical model of an important skip connection that has been extensively used in signal separation. For instance, in~\cite{jannson17} a model that uses skip connections based on ladder-like concatenations  improved performance when  skip-filtering connections were introduced. In addition to this, the skip-filtering connections are also an important ingredient in the Conv-TasNet model~\cite{tasnet-19} that has surpassed the oracle performance in speech signal separation.

Directions for future research include the linking between the demonstrated mapping functions and the spectral transportation matrices~\cite{flamary16_sot}. This could assist in a geometrical understanding of denoising functions in deep learning based signal separation. Transportation analysis would also allow the examination of latent information~\cite{sonoda17} that our work has not covered; however, the proposed method could be extended to that aim. Furthermore, examining another important type of skip connections that are commonly referred to as residual connections is also relevant. Residual connections play an important role in signal enhancement and denoising~\cite{joao_skip} and have been used in music source separation~\cite{takahashi17} and evaluation~\cite{uhl17}. Finally, expanding the proposed study to more advanced architectures is also relevant. That is because the current study has solely focused on data-dependent inter-frequency relationships, whereas it is well known that temporal information conveys much information in music signals. It should be stated though, that the mappings, with respect to the frequency structure, for recurrent and convolutional neural networks are not expected to deviate significantly from the mappings presented in this work. That is because the signal estimation is based on vector and matrix products as the models examined in this work.

%
%
\section*{Acknowledgments}
The research leading to these results has received funding from the European Union's H2020 Framework Programme (H2020-MSCA-ITN-2014) under grant agreement no 642685 MacSeNet. Stylianos Ioannis Mimilakis is supported by the by the German Research Foundation (AB 675/2-1, MU 2686/11-1). The authors would like to thank Rodrigo Pena (EPFL), Derry Fitzgerald (AudioSourceRE), Paul Magron (Tampere University), Luca Cuccovillo (Fraunhofer-IDMT), and the anonymous reviewers for the fruitful discussions and valuable feedback.


\begin{thebibliography}{10}
\bibitem{dl_wang_18}
D.~{Wang} and J.~{Chen}, ``Supervised speech separation based on deep learning:
  An overview,'' \emph{IEEE/ACM Transactions on Audio, Speech, and Language
  Processing}, vol.~26, no.~10, pp. 1702--1726, Oct 2018.

\bibitem{rafii18}
Z.~Rafii, A.~Liutkus, F.~R. St{\"o}ter, S.~I. Mimilakis, D.~FitzGerald, and
  B.~Pardo, ``An overview of lead and accompaniment separation in music,''
  \emph{IEEE/ACM Transactions on Audio, Speech, and Language Processing},
  vol.~26, no.~8, pp. 1307--1335, Aug 2018.

\bibitem{sisec18}
F.-R. St{\"o}ter, A.~Liutkus, and N.~Ito, ``{The 2018 Signal Separation
  Evaluation Campaign},'' in \emph{Proceedings of the {Latent Variable Analysis
  and Signal Separation: 14th International Conference on Latent Variable
  Analysis and Signal Separation}}, Surrey, United Kingdom, Jul. 2018.

\bibitem{vincent_08_den}
P.~Vincent, H.~Larochelle, Y.~Bengio, and P.-A. Manzagol, ``Extracting and
  composing robust features with denoising autoencoders,'' in \emph{Proceedings
  of the 25th International Conference on Machine Learning (ICML)}.\hskip 1em
  plus 0.5em minus 0.4em\relax New York, NY, USA: ACM, 2008, pp. 1096--1103.

\bibitem{vincent_den}
P.~Vincent, H.~Larochelle, I.~Lajoie, Y.~Bengio, and P.-A. Manzagol, ``Stacked
  denoising autoencoders: Learning useful representations in a deep network
  with a local denoising criterion,'' \emph{Journal of Machine Learning
  Research}, vol.~11, pp. 3371--3408, 2010.

\bibitem{grais17}
E.~M. Grais, G.~Roma, A.~J.~R. Simpson, and M.~D. Plumbley, ``Two-stage
  single-channel audio source separation using deep neural networks,''
  \emph{IEEE/ACM Transactions on Audio, Speech, and Language Processing},
  vol.~25, no.~9, pp. 1773--1783, Sept 2017.

\bibitem{williamson}
D.~Williamson, Y.~Wang, and D.~Wang, ``Complex ratio masking for monaural
  speech separation,'' \emph{IEEE/ACM Transactions on Audio, Speech, and
  Language Processing}, vol.~24, no.~3, pp. 483--492, March 2016.

\bibitem{mask_targets_speech}
Y.~Wang, A.~Narayanan, and D.~Wang, ``On training targets for supervised speech
  separation,'' \emph{IEEE/ACM Transactions on Audio, Speech, and Language
  Processing}, vol.~22, no.~12, pp. 1849--1858, Dec 2014.

\bibitem{uhl15}
S.~Uhlich, F.~Giron, and Y.~Mitsufuji, ``Deep neural network based instrument
  extraction from music,'' in \emph{Proceedings of the 40th International
  Conference on Acoustics, Speech and Signal Processing (ICASSP 2015)}, 2015,
  pp. 2135--2139.

\bibitem{grais16}
E.-M. Grais, G.~Roma, A.~Simpson, and M.-D. Plumbley, ``Single-channel audio
  source separation using deep neural network ensembles,'' in \emph{Proceedings
  of the 140th Audio Engineering Society Convention}, May 2016.

\bibitem{liutkus_alpha}
A.~Liutkus and R.~Badeau, ``Generalized {W}iener filtering with fractional
  power spectrograms,'' in \emph{Proceedings of the 40th International
  Conference on Acoustics, Speech and Signal Processing (ICASSP 2015)}, Apr.
  2015, pp. 266--270.

\bibitem{voran17}
S.~Voran, ``The selection of spectral magnitude exponents for separating two
  sources is dominated by phase distribution not magnitude distribution,'' in
  \emph{Proceedings of the 2017 IEEE Workshop on Applications of Signal
  Processing to Audio and Acoustics (WASPAA 2017)}, Oct. 2017.

\bibitem{fitz_masks}
D.~FitzGerald and R.~Jaiswal, ``On the use of masking filters in sound source
  separation,'' in \emph{Proceedings of the 15th International Conference on
  Digital Audio Effects (DAFx-12)}, Sep. 2012.

\bibitem{wening14}
F.~Weninger, J.~R. Hershey, J.~L. Roux, and B.~Schuller, ``Discriminatively
  trained recurrent neural networks for single-channel speech separation,'' in
  \emph{Proceedings of the 2014 IEEE Global Conference on Signal and
  Information Processing (GlobalSIP)}, Dec 2014, pp. 577--581.

\bibitem{huang}
P.-S. Huang, M.~Kim, M.~Hasegawa-Johnson, and P.~Smaragdis, ``Joint
  optimization of masks and deep recurrent neural networks for monaural source
  separation,'' \emph{IEEE/ACM Transactions on Audio, Speech, and Language
  Processing}, vol.~23, no.~12, pp. 2136--2147, Dec. 2015.

\bibitem{mim17_mlsp}
S.~I. Mimilakis, K.~Drossos, G.~Schuller, and T.~Virtanen, ``A recurrent
  encoder-decoder approach with skip-filtering connections for monaural singing
  voice separation,'' in \emph{Proceedings of the 27th IEEE International
  Workshop on Machine Learning for Signal Processing (MLSP)}, 2017.

\bibitem{jannson17}
A.~Jansson, E.~Humphrey, N.~Montecchio, R.~Bittner, A.~Kumar, and T.~Weyde,
  ``Singing voice separation with deep {U-Net} convolutional networks,'' in
  \emph{Proceedings of the 18th International Society for Music Information
  Retrieval Conference}, Suzhou, China, Oct. 2017.

\bibitem{mim17}
S.-I. Mimilakis, K.~Drossos, J.-F. Santos, G.~Schuller, T.~Virtanen, and
  Y.~Bengio, ``Monaural singing voice separation with skip-filtering
  connections and recurrent inference of time-frequency mask,'' in
  \emph{Proceedings of the 43rd International Conference on Acoustics, Speech
  and Signal Processing (ICASSP 2018)}, 2018.

\bibitem{drossos18}
K.~Drossos, S.~I. Mimilakis, D.~Serdyuk, G.~Schuller, T.~Virtanen, and
  Y.~Bengio, ``{MaD TwinNet}: Masker-denoiser architecture with twin networks
  for monaural sound source separation,'' in \emph{Proceedings of the 2018 IEEE
  International Joint Conference on Neural Networks (IJCNN)}, July 2018.

\bibitem{joint_pse}
J.~Lee, J.~Skoglund, T.~Shabestary, and H.~Kang, ``Phase-sensitive joint
  learning algorithms for deep learning-based speech enhancement,'' \emph{IEEE
  Signal Processing Letters}, vol.~25, no.~8, pp. 1276--1280, Aug 2018.

\bibitem{uhl17}
S.~Uhlich, M.~Porcu, F.~Giron, M.~Enenkl, T.~Kemp, N.~Takahashi, and
  Y.~Mitsufuji, ``Improving music source separation based on deep neural
  networks through data augmentation and network blending,'' in
  \emph{Proceedings of the 42nd International Conference on Acoustics, Speech
  and Signal Processing (ICASSP 2017)}, 2017, pp. 261--265.

\bibitem{takahashi17}
N.~Takahashi and Y.~Mitsufuji, ``Multi-scale multi-band densenets for audio
  source separation,'' in \emph{Proceedings of the 2017 IEEE Workshop on
  Applications of Signal Processing to Audio and Acoustics (WASPAA 2017)}, Oct.
  2017.

\bibitem{nug16}
A.~A. Nugraha, A.~Liutkus, and E.~Vincent, ``Multichannel music separation with
  deep neural networks,'' in \emph{Proceedings of the 24th European Signal
  Processing Conference (EUSIPCO)}, Aug 2016, pp. 1748--1752.

\bibitem{conservative_autoencoders}
D.-J. Im, M.~I. Belghazi, and R.~Memisevic, ``Conservativeness of untied
  auto-encoders,'' in \emph{Proceedings of the 30th AAAI Conference on
  Artificial Intelligence}, 2016.

\bibitem{den_ss}
J.~S\"{a}rel\"{a} and H.~Valpola, ``Denoising source separation,'' \emph{J.
  Mach. Learn. Res.}, vol.~6, pp. 233--272, Dec. 2005.

\bibitem{ps_masks}
H.~Erdogan, J.~R. Hershey, S.~Watanabe, and J.~L. Roux, ``Phase-sensitive and
  recognition-boosted speech separation using deep recurrent neural networks,''
  in \emph{Proceedings of the 40th International Conference on Acoustics,
  Speech and Signal Processing (ICASSP 2015)}, Apr. 2015, pp. 708--712.

\bibitem{LRP_dsp}
G.~Montavon, W.~Samek, and K.-R. M{\"u}ller, ``Methods for interpreting and
  understanding deep neural networks,'' \emph{Digital Signal Processing},
  vol.~73, pp. 1--15, 2018.

\bibitem{kolouri_ot_spmag}
S.~Kolouri, S.~R. Park, M.~Thorpe, D.~Slepcev, and G.~K. Rohde, ``Optimal mass
  transport: Signal processing and machine-learning applications,'' \emph{IEEE
  Signal Processing Magazine}, vol.~34, no.~4, pp. 43--59, July 2017.

\bibitem{flamary16_sot}
R.~Flamary, C.~F{\'e}votte, N.~Courty, and V.~Emiya, ``{Optimal spectral
  transportation with application to music transcription},'' in
  \emph{Proceedings of the 29th International Conference on Neural Information
  Processing Systems}, ser. NIPS'16, Barcelona, Spain, Dec. 2016.

\bibitem{hinton_distill}
G.~Hinton, O.~Vinyals, and J.~Dean, ``Distilling the knowledge in a neural
  network,'' in \emph{Proceedings of the 28th International Conference on
  Neural Information Processing Systems, Workshop on Deep Learning and
  Representation Learning}, ser. NIPS'15, 2015.

\bibitem{giannoulis11}
D.~Giannoulis, D.~Barchiesi, A.~Klapuri, and M.~D. Plumbley, ``On the
  disjointess of sources in music using different time-frequency
  representations,'' in \emph{Proceedings of the 2011 IEEE Workshop on
  Applications of Signal Processing to Audio and Acoustics (WASPAA 2011)}, Oct.
  2011, pp. 261--264.

\bibitem{kam14}
A.~Liutkus, D.~Fitzgerald, Z.~Rafii, B.~Pardo, and L.~Daudet, ``Kernel additive
  models for source separation,'' \emph{IEEE Transactions on Signal
  Processing}, vol.~62, no.~16, pp. 4298--4310, Aug 2014.

\bibitem{wang15}
Y.~Wang and D.~Wang, ``A deep neural network for time-domain signal
  reconstruction,'' in \emph{Proceedings of the 2015 IEEE International
  Conference on Acoustics, Speech and Signal Processing (ICASSP)}, April 2015,
  pp. 4390--4394.

\bibitem{nikkunen18}
G.~Naithani, J.~Nikunen, L.~Bramsl{{\o}}w, and T.~Virtanen, ``Deep neural
  network based speech separation optimizing an objective estimator of
  intelligibility for low latency applications,'' in \emph{Proceedings of the
  2018 IEEE International Workshop on Acoustic Signal Enhancement}, Sep. 2018.

\bibitem{wiener_matrix_guy}
W.~K. Pratt, ``Generalized {W}iener filtering computation techniques,''
  \emph{IEEE Transactions on Computers}, vol. C-21, no.~7, pp. 636--641, July
  1972.

\bibitem{mim_asilomar}
S.~I. Mimilakis, E.~Cano, D.~Fitzgerald, K.~Drossos, and G.~Schuller,
  ``Examining the perceptual effect of alternative objective functions for deep
  learning based music source separation,'' in \emph{Proceedings of the 52nd
  Asilomar Conference on Signals, Systems, and Computers}, 2018.

\bibitem{pascanu_montufar}
R.~Pascanu, G.~Montufar, and Y.~Bengio, ``On the number of response regions of
  deep feedforward networks with piecewise linear activations,'' in
  \emph{Proceedings of the International Conference on Learning Representations
  (ICLR’14)}, 2014.

\bibitem{papyan_ieee}
V.~Papyan, Y.~Romano, J.~Sulam, and M.~Elad, ``Theoretical foundations of deep
  learning via sparse representations: A multilayer sparse model and its
  connection to convolutional neural networks,'' \emph{IEEE Signal Processing
  Magazine}, vol.~35, no.~4, pp. 72--89, July 2018.

\bibitem{theodoridis_ml}
S.~Theodoridis, \emph{{Machine Learning: A Bayesian and Optimization
  Perspective}}, 1st~ed.\hskip 1em plus 0.5em minus 0.4em\relax Orlando, FL,
  USA: Academic Press, Inc., 2015.

\bibitem{wang16_datajacobian}
S.~Wang, A.-R. Mohamed, R.~Caruana, J.~Bilmes, M.~Plilipose, M.~Richardson,
  K.~Geras, G.~Urban, and O.~Aslan, ``Analysis of deep neural networks with the
  extended data jacobian matrix,'' in \emph{Proceedings of the 33rd
  International Conference on Machine Learning}, ser. ICML'16, vol.~48, 2016,
  pp. 718--727.

\bibitem{musdb18}
Z.~Rafii, A.~Liutkus, F.~St{\"o}ter, S.~I. Mimilakis, and R.~Bittner, ``The
  {MUSDB18} corpus for music separation,'' Dec 2017. [Online]. Available:
  \url{https://doi.org/10.5281/zenodo.1117372}

\bibitem{glorot}
X.~Glorot and Y.~Bengio, ``Understanding the difficulty of training deep
  feedforward neural networks,'' in \emph{Proceedings of the International
  Conference on Artificial Intelligence and Statistics (AISTATS'10)}, 2010, pp.
  249--256.

\bibitem{adam}
D.~P. Kingma and J.~Ba, ``Adam: {A} method for stochastic optimization,'' in
  \emph{Proceedings of the International Conference on Learning Representations
  (ICLR-15)}, 2015.

\bibitem{joao_skip}
J.~Santos and T.~Falk, ``Investigating the effect of residual and highway
  connections in speech enhancement models,'' in \emph{Proceedings of the 32nd
  International Conference on Neural Information Processing Systems, Workshop
  on Interpretability and Robustness in Audio, Speech, and Language}, ser.
  NIPS'18, Montreal, Canada, Dec. 2018.

\bibitem{dieleman14}
S.~{Dieleman} and B.~{Schrauwen}, ``End-to-end learning for music audio,'' in
  \emph{Proceedings of the 39th International Conference on Acoustics, Speech
  and Signal Processing (ICASSP 2014)}, May 2014, pp. 6964--6968.

\bibitem{tasnet-19}
Y.~{Luo} and N.~{Mesgarani}, ``{Conv-TasNet}: Surpassing ideal time–frequency
  magnitude masking for speech separation,'' \emph{IEEE/ACM Transactions on
  Audio, Speech, and Language Processing}, vol.~27, no.~8, pp. 1256--1266, Aug
  2019.

\bibitem{sonoda17}
S.~Sonoda and N.~Murata, ``Transportation analysis of denoising autoencoders: A
  novel method for analyzing deep neural networks,'' in \emph{Proceedings of
  the 2nd NeurIPS Workshop on Optimal Transport and Machine Learning (OTML
  2017)}, 2017.

\end{thebibliography}

\end{document}